\documentclass[prb,aps,english,twocolumn,notitlepage,superscriptaddress]{revtex4-2}

\usepackage{graphicx}% Include figure files
\usepackage{dcolumn}% Align table columns on decimal point
\usepackage{rotating}
\usepackage{lipsum}
\usepackage{xcolor}
\usepackage{bm}% bold math
\usepackage[T1]{fontenc}  
\usepackage{multirow,amsmath,amssymb}
 \usepackage{babel} 
 \usepackage{txfonts}
 \usepackage{epsfig,epsf,psfrag} 
\usepackage{graphicx} 
\usepackage{pslatex}
\usepackage{epic,eepic} 
\usepackage{color,pstcol}
\usepackage{pstricks} 
\usepackage{fancyhdr} 
\usepackage{tabularx}
\usepackage{physics}
\usepackage{url}
\usepackage{listings}
\usepackage{color} 
\usepackage{caption}
\usepackage{ulem}
\usepackage{balance}
\usepackage{subcaption}
\usepackage{float}
\usepackage[utf8]{inputenc}
\usepackage{comment}

\usepackage{hyperref}
\hypersetup{
 colorlinks=true,
 linkcolor=blue,
 anchorcolor = blue,
 citecolor = blue,
 filecolor = blue,
 urlcolor = blue
}
\usepackage{dsfont}

\graphicspath{{Images/}}
\def \be {\begin{equation}} 
\def \ee {\end{equation}}

\begin{document}
\title{$\Delta_T$ Noise in Mesoscopic Hybrid Junctions: Influence of Barrier Strength and Thermal Bias}
\author{Sachiraj Mishra}
\email{sachiraj29mishra@gmail.com}
\author{A Rajmohan Dora}
\email{arajmohan.dora@niser.ac.in}
\author{Tusaradri Mohapatra}
\email{tusaradri.mohapatra@niser.ac.in}
\author{Colin Benjamin}
\email{Corresponding Author, email:colin.nano@gmail.com}
\affiliation{School of Physical Sciences, National Institute of Science Education and Research, HBNI, Jatni-752050, India}
\affiliation{Homi Bhabha National Institute, Training School Complex, AnushaktiNagar, Mumbai, 400094, India }

\begin{abstract}
Quantum noise is a fundamental probe of quantum transport phenomena, offering insights into current correlations and wave-particle duality. A particularly intriguing form of such noise, $\Delta_T$ noise, emerges under a finite temperature difference in the absence of charge current at zero voltage bias. In this work, we investigate $\Delta_T$ noise in mesoscopic hybrid junctions incorporating insulating barriers, where the average charge current remains zero at zero bias. Using quantum shot noise measurements, we demonstrate that $\Delta_T$ noise in metal--insulator--superconductor (NIS) junctions is approximately 16 times greater than in metal--insulator--metal (NIN) counterparts. Our analysis further reveals that $\Delta_T$ noise exhibits a non-monotonic dependence on barrier strength---rising to a peak before declining---while increasing monotonically with the applied temperature bias. These findings underscore the rich interplay between thermal gradients and barrier properties in determining quantum noise characteristics in hybrid mesoscopic systems.
\end{abstract}
% \begin{abstract}
% {Quantum noise plays a pivotal role in understanding quantum transport phenomena, including current correlations and wave-particle duality. A recent focus in this domain is $\Delta_T$ noise, which arises due to a finite temperature difference in the absence of charge current at zero voltage bias. This paper investigates $\Delta_T$ noise in mesoscopic hybrid junctions with insulators, where the average charge current is zero at zero voltage bias, through the measurement of quantum shot noise, i.e., $\Delta_T$ noise. Notably, we find that the $\Delta_T$ noise in metal-insulator-superconductor junctions is {around 16 times} larger than in metal-insulator-metal junctions. Furthermore, our results reveal that $\Delta_T$ noise initially increases with barrier strength, peaks, and then decreases, while it shows a steady increase with temperature bias, highlighting the nuanced interplay between barrier characteristics and thermal gradients.}
% \end{abstract}

\maketitle

\section{Introduction}

$\Delta_T$ noise in a mesoscopic junction is calculated at a finite temperature gradient and at zero bias voltage with a vanishing charge current. At zero voltage bias, the average charge current in both normal metal--insulator--normal metal (NIN) and normal metal--insulator--superconductor (NIS) junctions vanishes even at finite temperature difference as electron-hole symmetry is preserved in these setups. However, these setups do generate extra noise apart from the obvious thermal noise, referred to as excess noise, which is the quantum shot noise contribution to the total quantum noise even at zero charge current and zero applied voltage bias. In Ref.~\cite{atomicscaleexpt}, this excess noise (or, quantum shot noise) is called $\Delta_T$ noise, because this noise is measured at zero average charge current. 

$\Delta_T$ noise has attracted considerable attention in recent research \cite{atomicscaleexpt, qcircuitexpt, tjunctionexpt, DeltaNS, generalbound, popoff, dTtheory, machinelearn,Prokudina2024} due to its relevance in thermal and quantum transport. It has been experimentally observed in atomic-scale molecular junctions \cite{atomicscaleexpt}, quantum circuits \cite{qcircuitexpt}, and in metallic tunnel junctions \cite{tjunctionexpt}, alongside growing theoretical interest \cite{DeltaNS, generalbound, popoff, dTtheory, machinelearn}. $\Delta_T$ noise is defined as the quantum shot noise at zero voltage bias due to a finite temperature gradient \cite{atomicscaleexpt}. Furthermore, $\Delta_T$ noise measurement is not restricted to a particular temperature range, allowing its application to conductors of varying sizes down to the atomic scale \cite{atomicscaleexpt, qcircuitexpt}. The versatility of $\Delta_T$ noise measurement makes it an effective tool for investigating and understanding the quantum transport in any conductor at mesoscopic scales. 

In this work, we study the zero voltage bias ($V=0$) transport regime leading to zero thermovoltage ($V_{th}=0$) but at a finite temperature bias ($\Delta T \neq 0$), wherein we calculate the $\Delta_T$ noise in a NIN and NIS junction. Our key findings reveal that $\Delta_T$ noise in NIN and NIS junctions are always positive, and that the $\Delta_T$ noise in a NIS junction is significantly enhanced compared to that in a NIN junction in the transparent limit. The study of ours underscores the non-monotonic dependence of $\Delta_T$ noise on barrier strength and its monotonic but quadratic dependence on temperature bias, emphasizing the intricate interplay between barrier characteristics and thermal gradients in NIN and NIS junctions.

The structure of this paper is as follows: ~\ref{theory} outlines the theoretical framework employed in our study, wherein ~\ref{theoryA} focuses on quantum transport in a NIN setup using the scattering approach. In ~\ref{theoryB}, we introduce quantum transport in the NIS setup. ~\ref{theoryC} presents the general theory of quantum noise, while ~\ref{theoryD} describes the method used to calculate the specific quantum noise and $\Delta_T$ noise in NIN and NIS junctions. In ~\ref{results}, we present our results on quantum noise, $\Delta_T$ noise, and normalized $\Delta_T$ noise in NIN and NIS junctions. Sec.~\ref{analysis} provides a comparative analysis of quantum noise, $\Delta_T$ noise, and normalized $\Delta_T$ noise in NIN and NIS junctions. We also elaborate on the fact why quantum noise in both NIN and NIS junctions remain almost unchanged with temperature bias although both $\Delta_T$ and normalized $\Delta_T$ noise vary quadratically with it in \ref{analysisA}. The paper concludes in ~\ref{conclusion} by summarizing the key findings and discussing potential experimental realizations. Derivations of the current, quantum noise and $\Delta_T$ noise in NIN and NIS junction{, analytical derivation of ratio of $\Delta_T$ noise are detailed in Appendices \ref{App_I}, \ref{App_ch_Qn}, \ref{App_ch_Dn}, and \ref{ratio} respectively.}

\section{Theory}
\label{theory}

\begin{figure}[h!]
\centering
\includegraphics[scale=0.295]{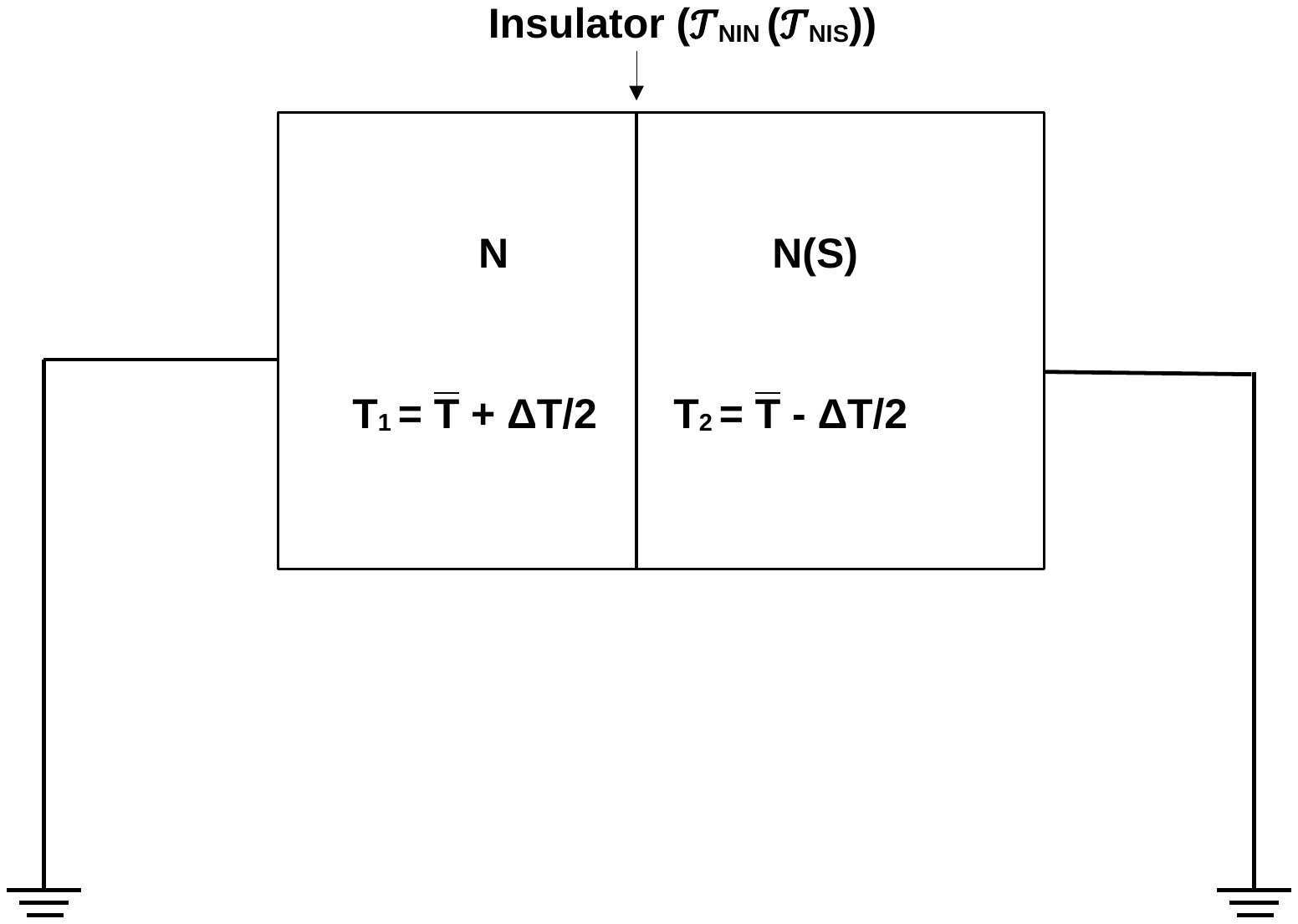}
\caption{{Schematic diagram of metal (N) - insulator (I) - metal (N) or metal-insulator-superconductor (NIS) junction. Both the left normal metal and right normal metal (or, superconductor) are grounded. $\mathcal{T}_{NIN}$ ($\mathcal{T}_{NIS}$) is the net transmission probability across the NIN (NIS) junction. $T_1 = \Bar{T} + \frac{\Delta T}{2}$ and $T_2 = \Bar{T} - \frac{\Delta T}{2}$ are the temperatures of the left and right normal metal (or, superconductor) respectively, where $\Bar{T}$ is the average temperature and $\Delta T$ is the temperature bias applied across the junction. For superconductor, the critical temperature $T_c > T_2$.}}
\label{fig1}
\end{figure}

%\end{widetext}

%\begin{figure}
 %    \centering
 %    \begin{subfigure}[b]{0.235\textwidth}
 %        \centering
 %        \includegraphics[width=\textwidth]{NIN_fig.png}
 %        \caption{NIN setup}
 %    \end{subfigure}
 %    \hspace{0.05cm}
 %    \begin{subfigure}[b]{0.235\textwidth}
 %        \centering
 %        \includegraphics[width=\textwidth]{NQN_fig.png}
 %        \caption{NQN setup}
 %    \end{subfigure}
 %       \caption{Schematics of a 1D different setups, (a) NIN junction, (b) NQN junction with QPC to measure quantum and $\Delta_T$ noise.  Temperature difference: $T_2-T_1$ $<$  average temperature: $(T_1 +T_2)/2$. }
 %        \label{fig1}
 %      \end{figure}

%{where, $E$ is the total energy calculated relative to $E_F$, i.e., $E = E' - E_F$ , $f = \left(1 + e^{E/k_B \bar{T}}\right)^{-1}$ is the equilibrium Fermi-Dirac distribution. 
 %Now, in our two-terminal setups such as NIN, NIS, we consider $V_1 = V_2 = 0$, while for NQN and NQS, we have $V_1 = V, V_2 = 0$ with $T_1 = \bar{T} -  \Delta T/2$ and $ T_2 = \bar{T} + \Delta T/2$, where $\bar{T}$ and $\Delta T$ are average temperature and temperature difference between reservoirs.}

\subsection{NIN junction}
\label{theoryA}
{In this section, we discuss the charge current in NIN junctions using {Landauer}-Buttiker formalism. }

{For a normal metal-insulator-normal metal setup in Fig. \ref{fig1}, where insulator is modeled by the potential $U \delta(x)$ at $x = 0$ and $U$ represents the barrier potential at the interface. The Hamiltonian for the NIN junction is,}
{
\begin{equation}
    \mathcal{H}_{NIN} = \begin{pmatrix}
        H_0(k) & 0 \\
        0 & -H_0^*(-k)
    \end{pmatrix},
\end{equation}}
{where {$H_0(k) = \frac{\hbar^2 k^2}{2m*} + U \delta(x) - E_F$}, with $m^{*}$ the effective electron mass, $k$ the wave vector, $E_{F}$ the Fermi energy, and $\hbar$ the reduced Planck’s constant ($\hbar = h/2\pi$). The wavefunction in both the left ($\psi_{N_1}$) and right normal metal ($\psi_{N_2}$) is given as}

{
\begin{equation}
\begin{split}
\psi_{N_1}(x) &= \left( \begin{array}{c}
1\\
0\\
\end{array} \right) (e^{i k_e x } + r e^{-i k_e x} ),~ \text{for} \hspace{0.08cm} x < 0,\\
\psi_{N_2}(x) &= \left( \begin{array}{c}
1\\
0\\
\end{array} \right) t e^{i k_e x},~ \text{for} \hspace{0.08cm} x \geq 0
\end{split}
\label{eq6}
\end{equation}}
{where $r$ and $t$ are the reflection and transmission amplitudes for the scatterer and the wave vector $k_e = \sqrt{\frac{2m^*}{\hbar^2}E'}$, where $E' = E_F + E$. We consider $E_F \gg E$, which means $k_e = k_F$, where $k_F = \sqrt{\frac{2m^*}{\hbar^2}E_F}$.
Using boundary conditions at the interface:}
{
\begin{equation}
\begin{split}
    \psi_{N_1}(x = 0) &= \psi_{N_2}(x = 0),\quad \text{and}\\
    \frac{d\psi_{N_2}(x = 0)}{dx} - \frac{d\psi_{N_1}(x = 0)}{dx} &= \frac{2 m^* U}{\hbar^2} \psi_{N_1}(x=0),
\end{split}
\label{eq7}
\end{equation}}
{Using Eq. (\ref{eq6}) into Eq. (\ref{eq7}), we get the reflection ($\mathcal{R}_{NIN}$) and transmission ($\mathcal{T}_{NIN}$) probabilities as a function of a dimensionless parameter $Z = m^* U/ (\hbar^2 k_F)$, as}
{
\begin{equation}
    \mathcal{R}_{NIN} = \frac{Z^2}{1 + Z^2}, \quad \mathcal{T}_{NIN} = \frac{1}{ 1 + Z^2}.
\label{eqn:TNIN}    
\end{equation}}

% $Z$ is the dimensionless barrier strength, $H$ is the barrier strength of the insulator, $m$ is the mass of the electron, $\hbar$ is the reduced Planck's constant, $k_F$ is the Fermi wave vector. Hence, $Z$ is independent of energy. Now, we can take $\mathcal{R}_{NIN}(-E)=\mathcal{R}_{NIN}(E)$.

% The scattering probabilities in NIN junction according to the BTK formalism are 
% \cite{BTK}, 
% \begin{equation}
%     \mathcal{R}_{NIN} = \frac{Z^2}{1 + Z^2}, \quad \mathcal{T}_{NIN} = \frac{1}{ 1 + Z^2}.
% \label{eqn:TNIN}    
% \end{equation}
{Here, $Z$ is a dimensionless and energy-independent function, which implies both $\mathcal{R}_{NIN}$ and $\mathcal{T}_{NIN}$ are energy-independent and obey $\mathcal{T}_{NIN}(-E)=\mathcal{T}_{NIN}(E)$ and $\mathcal{R}_{NIN}(-E)=\mathcal{R}_{NIN}(E)$.
From Landaeur-Buttiker formalism, the average charge current in a normal metal in a NIN junction \cite{BTK},}
{
\begin{equation}
\langle I^{NIN}_1 \rangle = \frac{2e}{h} \int^{\infty}_{-\infty}  \mathcal{T}_{NIN}  ( f_{1e} - f_{2e}) dE.
\label{eqn:ININ}
\end{equation}}

{The Fermi-Dirac distributions $f_{1e}$ and $f_{2e}$ as in Eq. (\ref{eqn:ININ}) can be expanded \cite{Benenti, atomicscaleexpt} in a linear expansion of temperature bias at zero voltage bias, where $T_1=\bar{T}+\Delta T/2$, and $T_2=\bar{T}-\Delta T/2$, are written as,}
{
\begin{equation}
\begin{split}
f_{1e}(E,k_B T_1) & \simeq f_0(E, k_B \bar{T}) + \frac{k_B \Delta T}{2} \frac{\partial f_0(E, k_B \bar{T})}{\partial (k_B \bar{T})} \\& \simeq f_0(E, k_B \bar{T}) - \frac{\Delta T}{2 \bar{T}} E \frac{\partial f_0(E, k_B \bar{T})}{\partial E},\\
f_{2e}(E,k_B T_2) & \simeq f_0(E, k_B \bar{T}) - \frac{k_B \Delta T}{2} \frac{\partial f_0(E, k_B \bar{T})}{\partial (k_B \bar{T})} \\& \simeq f_0(E, k_B \bar{T}) + \frac{\Delta T}{2 \bar{T}} E \frac{\partial f_0(E, k_B \bar{T})}{\partial E}, \\
\end{split}
\label{g1}
\end{equation}}

{where $f_0(E, k_B \bar{T})$ is the equilibrium Fermi-Dirac distribution at temperature $\bar{T}$, i.e., $f_0(E, k_B \bar{T}) = \frac{1}{1 + e^{\frac{E}{k_B \bar{T}}}}$. We have used the fact that $ \frac{\partial f_0(E, k_B \bar{T})}{\partial (k_B \bar{T})} = - \frac{E}{\bar{T}} \frac{\partial f_0(E, k_B \bar{T})}{\partial E}$ in Eq. (\ref{g1}). Thus, $f_{1e}(E,k_B T_1) -  f_{2e}(E,k_B T_2) = \frac{ \Delta T}{\bar{T}} E \frac{-\partial f_0(E, k_B \bar{T})}{\partial E}$, and $\langle I_1^{NIN} \rangle$ reduces to,}
{
\begin{equation}
\langle I^{NIN}_1 \rangle = \frac{2e}{h} \int^{\infty}_{-\infty}    dE \,\,  \mathcal{T}_{NIN}  \Bigg\{  E \left(-\frac{\partial f_0(E, k_B \bar{T})}{\partial E}\right) \Bigg\}  \frac{\Delta T}{\bar{T}}.
\label{eq1}
\end{equation}}

{Since $\mathcal{T}_{NIN}$ is energy independent, it can be taken out of the integral as in Eq. (\ref{eq1}) and the integration of $\int_{-\infty}^{\infty} E \left(\frac{-\partial f_0 (E, k_B \bar{T})}{\partial E}\right)$ is always zero as $\frac{-\partial f_0(E, k_B \bar{T})}{\partial E}$ is a symmetric function of $E$. Therefore, $\langle I_1^{NIN} \rangle$ vanishes at zero voltage bias and finite temperature bias. The vanishing integration is an indication that electron-hole symmetry is preserved. However, if $\mathcal{T}_{NIN}$ was an energy-dependent function, and an asymmetric function of $E$, then electron-hole symmetry would have been broken and then the integration would have been finite.
}

%\begin{widetext}
    
%\begin{figure}[h!]
%\centering
%\includegraphics[scale=0.45]{fig1.pdf}
%\includegraphics[scale=0.45]{fig2.pdf}
%\includegraphics[scale=0.45]{fig2.pdf}
%\caption{Transmission probability in (a) NIN, (b) NQN junction vs. electron energy ($E/E_F$), (c) mimicking a NQN junction with NIN junction replacing $Z$ with $ \sqrt{(1-\mathcal{T}_Q)/\mathcal{T}_Q}$.}
%\label{T_NIN}
%\end{figure}

%\end{widetext}

%{Using Eq. (\ref{eqn:IG_mat}), the average charge and heat current in terminal 1 in NIN setup can be written as}
%{
%\begin{equation}
%\begin{pmatrix}
%    \langle I_{1}^{NIN} \rangle \\
%    \langle J_{1}^{NIN} \rangle
%\end{pmatrix}
%=
%\begin{pmatrix}
%    G_{NIN} & S_{NIN} \\
%    P_{NIN} & \Pi_{NIN}
%\end{pmatrix}
%\begin{pmatrix}
%    V \\
%    -\Delta T
%\end{pmatrix},
%\end{equation}
%}

{Therefore, the symmetric nature of $\mathcal{T}_{NIN}$ and $\mathcal{R}_{NIN}$ refers to electron-hole symmetry. This symmetry of both $\mathcal{T}_{NIN}$ and $\mathcal{R}_{NIN}$ with respect to energy $E$ ensures that there is no net charge current ($\langle I_{NIN} \rangle = 0$) in the NIN junction at zero voltage bias. }

{
In the following section, We study the quantum transport property of NIS junction using the scattering approach.}

%{Now, with a QPC, the off-diagnonal Onsager coefficients are non-zero. In Fig. \ref{fig:I_NIN}(a), we plot $I^{NQN}$ as a function of $V$ for different values of $E_1$, i.e., $0.5 k_B \bar{T}$ and $1.0 k_B \bar{T}$. We observe that the current is zero at different voltage biases at these two different values of $E_1$. Similarly, in Fig. \ref{fig:G_NIN}(a), we plot $G^{NQN}$ as a function of $V$, where we observe that it is higher for $E_1 = 0.5 k_B \bar{T}$ compared to $E_1 = 1.0 k_B \bar{T}$. Similarly, in Fig. \ref{fig:J_NIN}(a), we plot $J^{NQN}$ as a function of $V$, we observe that here too, it vanishes at different values of $V$ for $E_1 = 0.5 k_B \bar{T}$ and $1.0 k_B \bar{T}$, whereas Fig. \ref{fig:K_NIN}(a) shows that $K^{NQN}$ is more for $E_1 = 0.5 k_B \bar{T}$ than $1.0 k_B \bar{T}$.}

%{In
%Fig. 10, we plot the $I^{NQS}$ and $G^{NQS}$ for different values of $E_1$. We observe that their magnitudes are almost double that of NQN or NQN case, whereas Fig. 11 shows that the magnitude of $\langle J_1^{NIS} \rangle$ and $K^{NIS}$ are almost double of NQN or NQN case.}

\subsection{NIS junction}
\label{theoryB}
{In this section, we consider a normal metal-insulator-superconductor junction, where just like NIN, the insulator is modelled by the potential $U \delta(x)$ at $x = 0$. }
The Hamiltonian for a NIS junction, using the Bogoliubov-de Gennes (BDG) formalism as described in Ref. ~\cite{BTK}, can be expressed as follows:

\begin{equation}
\mathcal{H}=\left(
\begin{array}{cc}
H_0(\textbf{k}) & \Delta_0 \Theta(x) \\
\Delta_0^{\dagger} \Theta(x) & -H^*_0(-\textbf{k})
\end{array}
\right),
\label{Eq:H}
\end{equation}
with $\Theta(x)$ is Heaviside theta function, $\Delta_0$ is the superconducting gap ~\cite{Tc}, and $k_B$ is Boltzmann constant. $H_0 (\textbf{k}) = -\frac{\hbar^2 \textbf{k}^2}{ 2 m^2} + U \delta(x) -E_F $, where $U$ represents the barrier potential at the interface, $E_F$ is the Fermi energy, and $m^*$ is the mass of the electron.

%\begin{figure}[h!]
%\centering
%\includegraphics[scale=0.65]{FIG1_new.png}
%\caption{Schematic of 1D normal metal/insulator/superconductor junction (NIS) with a quantum point contact (QPC) located at the interface of NIS junction ($x=0$).}
%\label{fig:NIS}
%\end{figure}

%\begin{figure}
%     \centering
%     \begin{subfigure}[b]{0.235\textwidth}
%         \centering
%         \includegraphics[width=\textwidth]{NIS_fig.png}
%         \caption{NIN setup}
%     \end{subfigure}
%     \hspace{0.05cm}
%     \begin{subfigure}[b]{0.235\textwidth}
%         \centering
%         \includegraphics[width=\textwidth]{NQS_fig.png}
%         \caption{NQN setup}
%     \end{subfigure}
%        \caption{Schematics of a 1D different setups, (a) NIS junction, (b) NQS junction with QPC to measure quantum and $\Delta_T$ noise.  Temperature difference: $T_2-T_1$ $<$  average temperature: $(T_1 +T_2)/2$. }
%         \label{fig1}
%       \end{figure}

%\begin{widetext}

%\end{widetext}

The wave functions in normal metal ($N$) and superconductor ($S$) regions, for an electron incident from normal metal $N$ are given as (see, Fig.~\ref{fig1}),
\begin{align}
\psi_{N}(x) &= \left( \begin{array}{c}
1\\
0\\
\end{array} \right) (e^{i k_e x } + b e^{-i k_e x} ) +
a \left( \begin{array}{c}
0\\
1\\
\end{array} \right) e^{i k_h x} ,~ \text{for} \hspace{0.08cm} x < 0, \nonumber \\
\psi_{S}(x) &= c \left( \begin{array}{c}
u\\
v
\end{array} \right) e^{i k^s_e x} +
d \left( \begin{array}{c}
v\\
u
\end{array} \right) e^{-i k^s_h x} ,~\text{ for} \hspace{0.09cm} x > 0,
\label{eqn:wf}
\end{align}
where $k_{e,h}$ are wave-vectors in normal metal for electron and hole, i.e., $k_{e,h}= \sqrt{ \frac{2 m^*}{ \hbar^2} ( E_F \pm E )}$, wherein $E$ is the excitation energy of the electron and $k^s_{e,h}$ are the wave-vectors in the superconductor for electron-like and hole-like quasiparticles, i.e., $k^s_{e,h} = \sqrt{ \frac{2 m^*}{ \hbar^2} ( E_F \pm \sqrt[]{E^2-\Delta_0^2} ) }$. The coherence factors for energy $E$ above the superconducting gap $\Delta_0$ are $u(v) = \left[ \frac{1}{2} \left\{ 1 \pm \frac{ \sqrt[]{ E^2 - \Delta^2_0 }}{E} \right\} \right]^{1/2} $. {When the Fermi energy is significantly larger than the energy of the incident electron or hole and the superconducting gap, i.e., $ E_F \gg E, \Delta_0 $, the wavevectors of all electrons and holes effectively become the Fermi wavevector under the Andreev approximation \cite{BTK}, such as $k_e=k_h=k^s_e=k^s_h=k_F$. All particle energies in both contacts in a NIS junction are measured relative to the Fermi energy \cite{BTK, datta}. } In a NIS junction, an incident electron on the interface undergoes two possible reflections: normal reflection, where it is reflected as an electron, or Andreev reflection, where it is reflected as a hole. The Andreev reflection is characterized by the amplitude $a$, while the normal reflection is denoted by the amplitude $b$. Furthermore, there are two transmission amplitudes: $c$ represents the transmission amplitude of an electron transmitted as an electron, and $d$ represents the transmission amplitude of an electron transmitted as a hole, and their respective probabilities are defined as $\mathcal{A} =  |a|^2$, $\mathcal{B}= |b|^2$, $\mathcal{C} = (|u|^2 -|v|^2) |c|^2$ and $\mathcal{D}= (|u|^2 -|v|^2) |d|^2$ \cite{BTK}. The prefactor $(|u|^2 -|v|^2)$ arises from the need to conserve the probability current, see Ref. \cite{BTK}.
% The scattering amplitude $s^{\rho \gamma}_{i k}$ describes the behavior of a particle with type $\gamma$ (either electron $e$ or hole $h$) incident from contact $k$ (either $N$ or $S$) that is either reflected or transmitted to contact $i$ (either $N$ or $S$) as a particle of type $\rho$ (either electron $e$ or hole $h$).

The boundary conditions for electron incident from $N$, at the interface $x=0$ (see, Fig~\ref{fig1}) in a NIS junction are,
\begin{eqnarray}
\psi_{N}(x=0) &=& \psi_{S}(x=0) , \nonumber \\
\frac{d \psi_{S} (x=0)}{dx} - \frac{d \psi_{N}(x=0)}{dx} &=& \frac{2 m^* U}{\hbar^2} \psi_{N} (x=0).
\label{BC}
\end{eqnarray}
Incorporating Eq.~(\ref{eqn:wf}) into Eq.~(\ref{BC}), we get scattering amplitudes with barrier strength characterized by dimensionless parameter $Z$. 
% Now, the transmission probabilities $\mathcal{A}$ and $\mathcal{B}$ are given as

%\begin{figure}[h!]
%\centering
%\includegraphics[scale=0.64]{fig3.pdf}
%\caption{Transmission probability via NIS junction.}
%\label{fig:I_NIS}

% Add labels and horizontal left-right arrow
%\begin{picture}(0,0)
%    \put(-20, 130){\makebox(0,0)[c]{\textbf{$\mathcal{T}_{NIS}$}}} % Label for T_Q
%    \put(70, 30){\makebox(0,0)[c]{\textbf{$E/\Delta_0$}}} % Label for E

    % Custom long arrow using pict2e
%    \put(85, 30){\makebox(0,0)[l]{\textbf{\vector(1,0){20}}}} % Long right arrow
%    \put(-20, 150){\makebox(0,0)[c]{\textbf{\vector(0,1){20}}}} % Long up arrow
%\end{picture}
%\end{figure}

%\begin{figure}[h!]
%\centering
%\includegraphics[scale=0.64]{fig4.pdf}
%\caption{Transmission probability via NQS junction.}
%\label{fig:I_NIS}

% Add labels and horizontal left-right arrow
%\begin{picture}(0,0)
%    \put(-20, 130){\makebox(0,0)[c]{\textbf{$\mathcal{T}_{NQS}$}}} % Label for T_Q
%    \put(70, 30){\makebox(0,0)[c]{\textbf{$E/\Delta_0$}}} % Label for E

    % Custom long arrow using pict2e
%    \put(85, 30){\makebox(0,0)[l]{\textbf{\vector(1,0){20}}}} % Long right arrow
%    \put(-20, 150){\makebox(0,0)[c]{\textbf{\vector(0,1){20}}}} % Long up arrow
%\end{picture}
%\end{figure}

{Once we get the scattering probabilities $\mathcal{A}, \mathcal{B}, \mathcal{C}$ and $\mathcal{D}$, we can find the charge current in the NIS junction. The net transmission probability via the NIS junction is $\mathcal{T}_{NIS}$ and it is equal to $1 + \mathcal{A} - \mathcal{B}$ \cite{BTK}.} {The scattering probabilities in the NIS junction from the BTK formalism \cite{BTK},
$\mathcal{A} = \frac{u_0^2 v_0^2}{\gamma^2},\quad
\mathcal{B} = \frac{(u_0^2 - v_0^2)^2 Z^2 (1+Z^2)}{\gamma^2},$
where,
$\gamma^2 = \big[ u_0^2 + Z^2 (u_0^2 - v_0^2) \big]^2,
u_0^2 = 1 - v_0^2 
= \tfrac{1}{2}\left\{ 1 + \sqrt{1 - \frac{\Delta^2}{E^2}} \right\}$. $\mathcal{A}$ and $\mathcal{B}$ depends on $E^2$. Hence, they are even functions of energy $E$, i.e., $\mathcal{A}(-E)=\mathcal{A}(E)$ and $\mathcal{B}(-E)=\mathcal{B}(E)$.
From Landaeur-Buttiker formalism, the average charge current in a normal metal in a NIS junction as derived in Sec. \ref{App_I},} 
{
\begin{equation}
\begin{split}
\langle I^{NIS}_1 \rangle &=\frac{2e}{h} \int^{\infty}_{-\infty}  (1+\mathcal{A}-\mathcal{B})  ( f_{1e} - f_{2e}) dE\\& = \frac{2e}{h} \int^{\infty}_{-\infty}  \mathcal{T}_{NIS}  ( f_{1e} - f_{2e}) dE,
\end{split}
\label{eqn:I_Hir}
\end{equation}}
{where, $\mathcal{T}_{NIS}=1+\mathcal{A}-\mathcal{B}$. Since $\mathcal{A}, \mathcal{B}$ are even function of energy, then $\mathcal{T}_{NIS}(-E)=\mathcal{T}_{NIS}(E)$. Using the expansion as shown in Eq. (\ref{g1}), we get,}

{
\begin{equation}
\langle I^{NIS}_1 \rangle = \frac{2e}{h} \int^{\infty}_{-\infty}    dE \,\,  \mathcal{T}_{NIS}  \Bigg\{  E \left(-\frac{\partial f_0(E, k_B \bar{T})}{\partial E}\right) \Bigg\}  \frac{\Delta T}{\bar{T}}. 
\end{equation}}

{Since $\mathcal{T}_{NIS}$ is an even function of energy, the integral 
$\int_{-\infty}^{\infty} E \, \mathcal{T}_{NIS} \, \frac{\partial f_0(E, k_B \bar{T})}{\partial E} \, dE$ vanishes. 
This is because $\frac{\partial f_0(E, k_B \bar{T})}{\partial E}$ is symmetric in $E$, $\mathcal{T}_{NIS}$ is also symmetric, 
and the factor of $E$ makes the integrand overall antisymmetric. Consequently, $\langle I_1^{NIS} \rangle$ is zero 
at zero voltage bias under a finite temperature gradient, indicates that electron-hole symmetry is preserved. }

Next, we calculate the quantum noise and $\Delta_T$ noise in two setups: NIN and NIS.

{}

\subsection{Quantum noise}
\label{theoryC}
{Quantum noise auto (cross) correlation defines the charge current-current correlation in the normal metal $N$ in the NIN junction. In general, quantum noise correlations \cite{noise, thermalnoise, datta, martin} between current in contacts $p$ and $q$ at time $t$ and $t^{\prime}$ can be written as $Q_{p q}(t-t') \equiv \langle \Delta I_{p}(t) \Delta I_{q}(t') + \Delta I_{q}(t') \Delta I_{p}(t) \rangle $, where $\Delta I_{p}(t)= I_{p}(t) - \langle I_{p}(t) \rangle$. Fourier transforming the time dependent quantum noise yields the frequency dependent quantum noise power, which is expressed as $ \delta(\omega + \bar{\omega}) Q_{pq}(\omega) \equiv \frac{1}{2 \pi } \langle \Delta I_{p}(\omega) \Delta I_{q}(\bar{\omega}) + \Delta I_{q}(\bar{\omega}) \Delta I_{p}(\omega) \rangle $. The expression for zero frequency quantum noise auto-correlation for a NIN junction \cite{datta, martin} is,}

\begin{widetext}
{
\begin{eqnarray}
Q^{NIN}_{11} (\omega =0) &=& \frac{2e^2}{h} 
\int \sum_{\substack{k,l \in \{1, 2\}}}  A_{k;l}(1 ,E) A_{l;k}(1 ,E) \nonumber 
  \times (\textit{f}_{ke}(E) [1-\textit{f}_{le}(E)]+\textit{f}_{le}(E) [1-\textit{f}_{ke}(E)]) dE, \\
 &=& \frac{4 e^2}{h} \biggl[ 
\int^{\infty}_{-\infty} F^{NIN}_{11 th} 
\bigl\{ f_{1e} (1- f_{1e}) + f_{2e} (1- f_{2e}) \bigr\} dE + \int^{\infty}_{-\infty} F^{NIN}_{11 sh} (f_{1e} - f_{2e})^2 dE \biggr] = Q^{NIN}_{11 th} + Q^{NIN}_{11 sh},
\label{eqn:S_NIN}
\end{eqnarray}}
\end{widetext}
{where $f_{ke(le)}(E) = \left[ 1+ e^{\frac{E - e V_{k(l)}}{k_B T_{k(l)}}} \right]^{-1}$ is the Fermi-Dirac distribution in terminal $k$ or $l$. {$f_{1(2)} = \left[1 + e^{\frac{E - e V_{1(2)}}{k_B T_{1(2)}}}\right]^{-1}$} is the Fermi-Dirac distribution in terminal 1 or 2. For the NIN junction, terminals 1 and 2 are normal metals, and for the NIS junction, terminal 1 is normal metal, and terminal 2 is a superconductor. The derivation of $F_{11th}^{NIN}  =  \mathcal{T}_{NIN}$ and $F_{11sh}^{NIN} = \mathcal{T}_{NIN}(1 - \mathcal{T}_{NIN})$ is done in Appendix \ref{App_ch_Qn}. In Eq. (\ref{eqn:S_NIN}), $Q_{11th}^{NIN}$ is the thermal noise-like contribution and $Q_{11sh}^{NIN}$ is the shot noise-like contribution. {The extra term $2$ in Eq. (\ref{eqn:S_NIN}) comes as a result of spin-degeneracy.}}

{Similarly, one can also calculate the quantum noise in NIS junctions. In a NIS junction, quantum noise correlations \cite{datta, martin} between current in contacts $p$ and $q$ at time $t$ and $t^{\prime}$ can be written as $Q^{xy}_{p q}(t-t') \equiv \langle \Delta I^x_{p}(t) \Delta I^y_{q}(t') + \Delta I^y_{q}(t') \Delta I^x_{p}(t) \rangle $, where $\Delta I^x_{p}(t)= I^x_{p}(t) - \langle I^x_{p}(t) \rangle$, and $\{x,y\} \in \{ e,h\}$. Fourier transforming time dependent quantum noise yields the frequency dependent quantum noise power, which is expressed as $ \delta(\omega + \bar{\omega}) Q^{xy}_{pq}(\omega) \equiv \frac{1}{2 \pi } \langle \Delta I^x_{p}(\omega) \Delta I^y_{q}(\bar{\omega}) + \Delta I^y_{q}(\bar{\omega}) \Delta I^x_{p}(\omega) \rangle $. The expression for zero frequency quantum noise auto-correlation for a two terminal NIS junction \cite{datta, martin} is,}
\begin{widetext}
\begin{align}
{Q^{NIS}_{11} (\omega =0)} &= {\sum_{x, y \in \{e, h \}} Q_{11}^{xy}} \nonumber = \frac{2e^2}{h} 
\int \sum_{\substack{k,l \in \{1, 2\}, \\ x,y,\gamma,\eta \in \{e,h\}}} 
sgn(x) sgn(y) A_{k,\gamma;l,\eta}(i x,E) A_{l,\eta;k,\gamma}(j y,E) \nonumber  \times(\textit{f}_{k \gamma} [1-\textit{f}_{l \eta}] + \textit{f}_{l \eta} [1-\textit{f}_{k \gamma}]) dE \nonumber \\
&= Q^{NIS}_{11 th} + Q^{NIS}_{11 sh}, \,\,\, \text{where} \,\,Q_{11th}^{NIS} = \frac{4e^2}{h} \int^{\infty}_{-\infty} \biggl[ (1 - \mathcal{B} + 3 \mathcal{A} ) f_{1e} (1-f_{1e}) + (1 - \mathcal{B} - \mathcal{A})~ f_{2e} (1-f_{2e}) \biggl] dE,\,\,\, \text{and} \nonumber \\
& \quad \quad \quad \quad \quad \quad \quad \quad \quad \quad  Q_{11sh}^{NIS} = \frac{4e^2}{h} \int^{\infty}_{-\infty} \biggl[ \mathcal{A}(1-\mathcal{A}) + \mathcal{B}(1-\mathcal{B}) + 2 \mathcal{A} \mathcal{B}  \biggr] ( f_{1e} -f_{2e})^2 dE,
\label{eqn:S_NIS}
\end{align}

\end{widetext}

{where, $f_{k \gamma}(E) = \left[ 1+ e^{\frac{E+ sgn(\gamma) e V_{k}}{k_B T_{k}}} \right]^{-1}$ is the Fermi-Dirac distribution in terminal $k$ for particle type $\gamma \in \{e, h \}$. $sgn(x)= sgn(y) = +1$ for electron and $-1$ for hole. {The extra term $2$ in Eq. (\ref{eqn:S_NIS}) comes as a result of spin-degeneracy. The detailed derivation of Eq. (\ref{eqn:S_NIS}) is given in Appendix \ref{App_ch_Qn}}. Here, $Q_{11th}^{NIS}$ is the thermal noise-like contribution, whereas $Q_{11sh}^{NIS}$ is the shot noise-like contribution. }

{In the next sub-section, we discuss the method to calculate $\Delta_T$ noise in NIN and NIS junctions in detail.} 

\subsection{$\Delta_T$ noise in NIN/NIS junctions}
\label{theoryD}

{To calculate the $\Delta_T$ noise in the NIN junction, i.e., $\Delta_T^{NIN}$, we use $Q_{11sh}^{NIN}$ from Eq. (\ref{eqn:S_NIN}).} {In this work, we define the temperature of the normal metal
as $T_1 = \bar{T} + \frac{\Delta T}{2}$ and the temperature of the superconductor as $T_2 = \bar{T} - \frac{\Delta T}{2}$ at zero voltage bias ($V = 0$), where $\bar{T}$ represents the average temperature of
the system and $\Delta T$ denotes the temperature bias. We specifically consider the regime where $\Delta T \ll \bar{T}$, i.e., the linear response regime.} {For this case, the average charge current, denoted as $\langle I_1^{NIN} \rangle$, is zero due to electron-hole symmetry even in the presence of finite temperature bias. In the limit, $E \ll E_F$, for which $\mathcal{T}_{NIN}$ is independent of energy.  Therefore, from Eq. (\ref{eqn:S_NIN}), the $\Delta_T^{NIN}$ noise is given as,}
{
\begin{equation}
    \Delta_T^{NIN} = \frac{4e^2}{h} F_{11sh}^{NIN}  \int_{-\infty}^{\infty} dE  (f_{1e} - f_{2e})^2.
\label{eq15}    
\end{equation}}

  {where, $F_{11sh}^{NIN} = \mathcal{T}_{NIN} (1 - \mathcal{T}_{NIN})$. We further derive the expression of $\Delta_T^{NIN}$ in detail in Appendix \ref{App_ch_Dn}, which gives the} $\Delta_T^{NIN}$ to be
 {
\begin{equation}
    \Delta_T^{NIN} = \frac{4e^2}{h} F_{11sh}^{NIN}k_B \bar{T} \left(\frac{\pi^2}{18} - \frac{1}{3}\right) \frac{\Delta T^2}{\bar{T}^2}.
\label{eq300}
\end{equation}}

{Eq. (\ref{eq300}) is the exact expression for $\Delta_T$ noise in NIN junction at zero voltage bias and finite temperature bias. Similarly, to calculate the $\Delta_T$ noise in the NIS junction, we use Eq. (\ref{eqn:S_NIS}) with the same voltage and temperature biases, i.e., \(V = 0\) and \(T_1 = \bar{T} + \Delta T / 2\) and \(T_2 = \bar{T} - \Delta T / 2\). Here too, the charge current \(\langle I_1^{NIS} \rangle\) is zero, and the $\Delta_T$ noise is effectively the quantum shot noise at zero voltage bias, represented as \(\Delta_{T}^{NIS}\).}

{Using Eq. (\ref{eqn:S_NIS}), $\Delta_T^{NIS}$ is given as,}
{
\begin{equation}
    \Delta_T^{NIS} = \frac{4e^2}{h} \int_{-\infty}^{\infty}dE F_{11sh}^{NIS}(f_{1e} - f_{2e})^2 dE,
    \label{eq100}
\end{equation}}

{where, $F_{11sh}^{NIS} = \mathcal{A} (1 - \mathcal{A}) + \mathcal{B} (1 - \mathcal{B}) + 2 \mathcal{A} \mathcal{B}$.} {Doing the Taylor series expansion of $(f_{1e} - f_{2e})^2$ similar to Eq. (\ref{eq17}) of Appendix \ref{App_ch_Dn}, we get,}
{
\begin{equation}
    \Delta_T^{NIS} = \frac{4e^2}{h} \frac{(\Delta T)^2}{k_B^2 \bar{T}^4}  \int_{-\infty}^{\infty} dE F_{11sh}^{NIS} \frac{E^2 e^{\frac{2E}{k_B \bar{T}}}}{\left(1 + e^{\frac{E}{k_B \bar{T}}}\right)^4}.
    \label{eq101}
\end{equation}}

{Unlike NIN junction, wherein exact analytical expression for $\Delta_T^{NIN}$ in Eq. (\ref{eq300}) can be derived, it is difficult to derive an exact analytical expression for $\Delta_T^{NIS}$ since {$\mathcal{A}$} and {$\mathcal{B}$} in Eq. (\ref{eq101}) are energy dependent, see Sec. \ref{theory} B. Therefore, we evaluate the integral in Eq. (\ref{eq101}) numerically for specific $Z$ values: 0 (transparent limit), 1 (intermediate limit) and 5 (tunnel limit) with average temperature $\bar{T}$ $ = 3K$.}

{In the next section, we present the detailed results for $\Delta_T$ noise in NIN and NIS junctions. }

\section{Results and Discussion}
\label{results}

{This section delves into the main findings of our work. Here, we discuss both quantum noise and $\Delta_T$ noise in NIS and NIN junctions at zero applied voltage bias and finite temperature bias. Therefore, the charge quantum noise is measured at zero charge current. The charge quantum noise measured at zero charge current is approximately same as its thermal noise component, because it is 100 times larger than its shot noise-like component, i.e., the $\Delta_T$ noise. These results have been produced in Mathematica, the Mathematica notebook can be downloaded in Ref. \cite{github}.}

{We analyze the $\Delta_T$ noise and quantum noise in both NIN and NIS junctions across three different transport regimes: the transparent limit, the intermediate limit, and the tunneling limit, as illustrated in Figs. \ref{fig100} and \ref{fig101}. The transparent limit corresponds to $Z \to 0$, the intermediate limit to $Z = 1$, and the tunneling limit to $Z = 5$. Additionally, we focus on the linear response regime, where the temperature bias $\Delta T$ is much smaller than the average temperature $\bar{T}$. Specifically, we set $\bar{T} = 3K$ and vary $\Delta T$ from $0K$ to $0.5K$, ensuring that our analysis remains within the linear response framework. These conditions are depicted in Figs. \ref{fig100} and \ref{fig101}}. In Fig.~\ref{fig100}, we present the total $\Delta_T$ noise in NIN junction, i.e., $\Delta_T^{NIN}$, as a function of both interface barrier strength $Z$ and temperature bias $\Delta T$. From Fig.~\ref{fig100}(a), it is clear that $\Delta_T^{NIN}$ exhibits a non-monotonic dependence on $Z$ with a peak at $Z \to 1$ (intermediate barrier strength). {However, in the tunneling limit ($Z \to 5$), the $\Delta_T$ noise decreases.} The inset of Fig.~\ref{fig100}(a) shows the behavior of $Q_{11}^{NIN}$, which follows a contrasting trend. $Q_{11}^{NIN}$ achieves its peak in the transparent limit ($Z \to 0$). As the barrier strength $Z$ increases, $Q_{11}^{NIN}$ steadily declines, eventually vanishing in the tunneling regime. This behavior illustrates the suppression of quantum noise with increasing $Z$. {We observe that the quantum noise $Q_{11}^{NIN}$ reaches its maximum in the transparent limit ($Z \to 0$) and decreases as $Z$ increases. This behavior arises because the shot noise-like contribution at zero voltage bias and finite temperature bias with zero average charge current (represented by $\Delta_T$ noise in our case) is much smaller compared to the thermal noise-like contribution. Since the thermal noise-like contribution carries the same information as charge conductance, the total quantum noise is approximately equal to the thermal noise-like contribution, leading to the same overall trend as the charge conductance.}

{We further investigate the dependence of $\Delta_{T}^{NIN}$ on temperature bias $\Delta T$ for fixed values of {$Z \to 0$ (transparent), $Z = 1$ (intermediate), and $Z = 5$ (tunneling)}, as presented in Fig.~\ref{fig100}(b). It is very small for {$Z \to 0$} and increases quadratically with $\Delta T$ for $Z = 1$ and $Z = 5$, as seen in the inset of Fig.~\ref{fig100}(b). In contrast, for {$Z \to 0$}, $Q_{11}^{NIN}$ attains its maximum value and progressively decreases for $Z = 1$ and $Z = 5$, as shown in the inset of Fig. \ref{fig100}(b). Remarkably, $Q_{11}^{NIN}$ displays a nearly constant variation with $\Delta T$ across all $Z$ values. {These findings highlight the distinct dependencies of quantum noise and $\Delta_T$ noise on barrier strengths and thermal gradients in an NIN junction.}}

\begin{widetext}

\begin{figure}[h!]
\includegraphics[scale=.38]{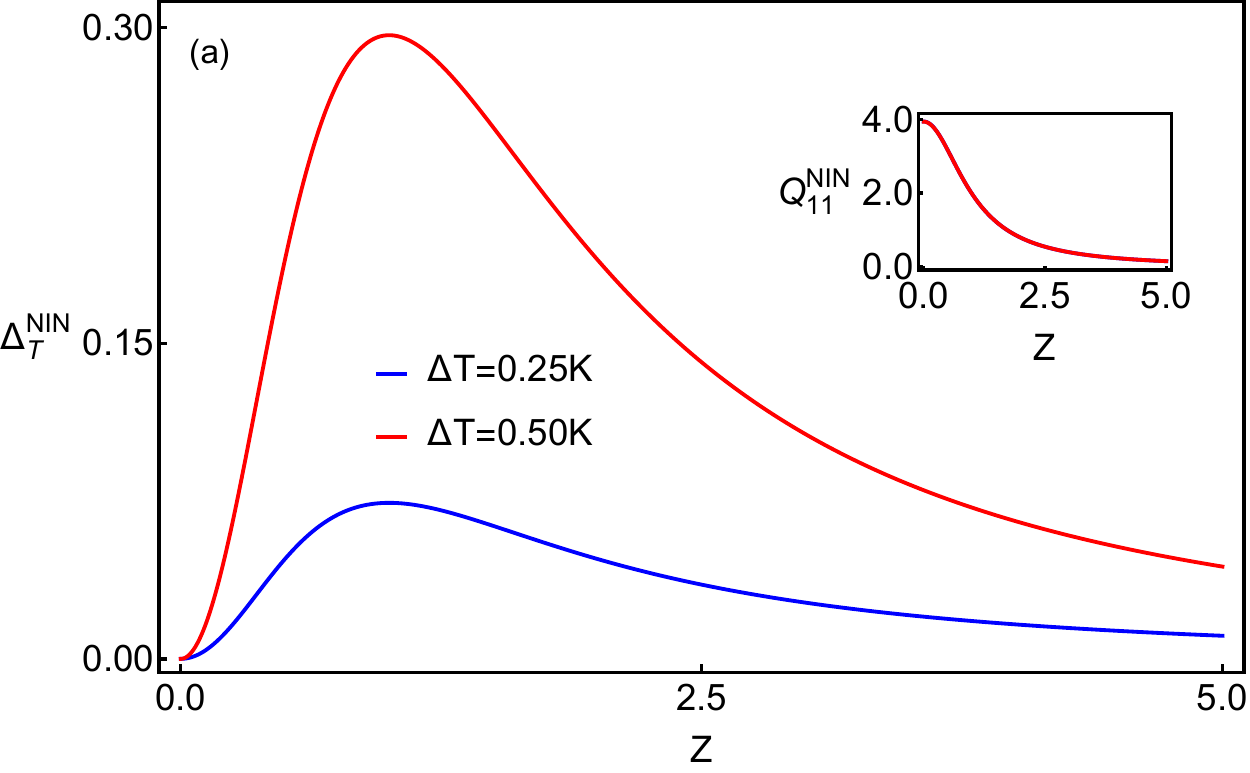}
\includegraphics[scale=.38]{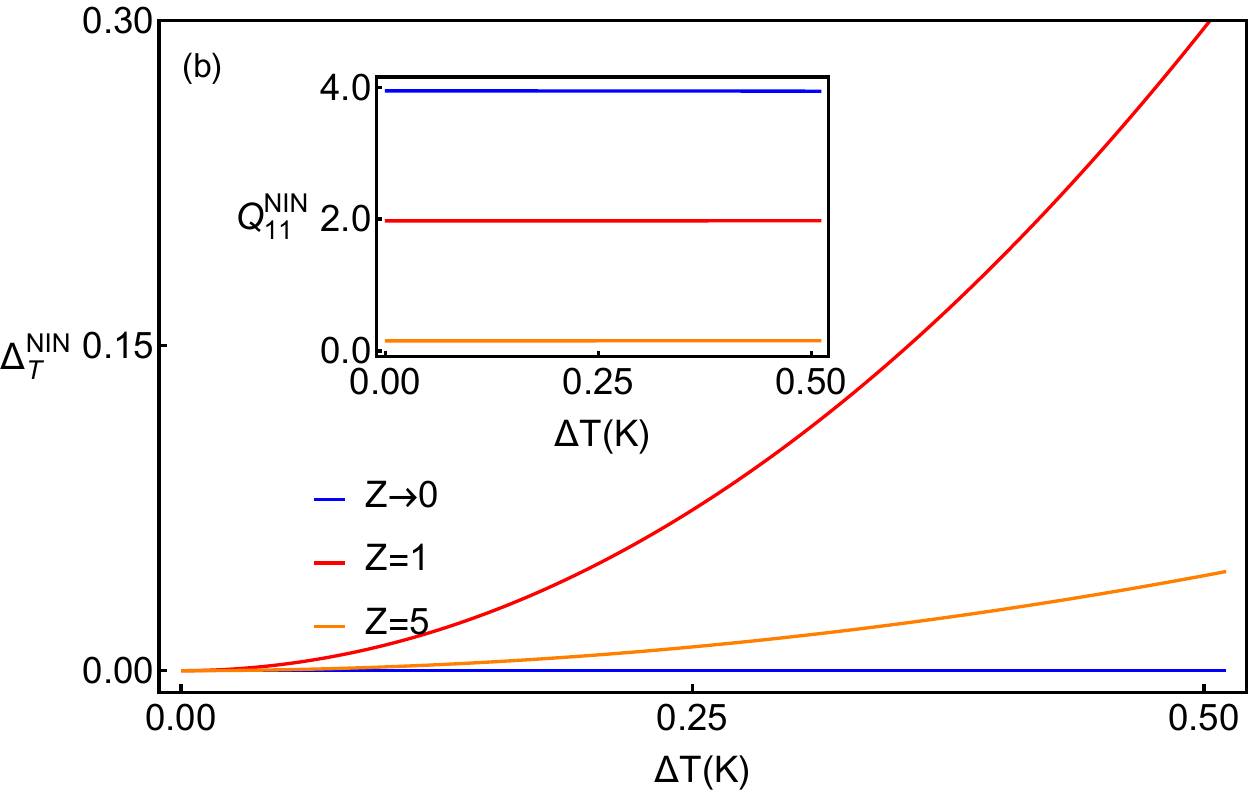}
\caption{$\Delta_T^{NIN}$ noise (Quantum noise $Q^{NIN}_{11}$ in units of $\frac{2e^2}{h} k_B \bar{T}$ in inset) in units of $\frac{2e^2}{h} 10^{-2} k_B \bar{T}$ (a)  vs. $Z$, with $\Delta T=0.25K$ (blue), $\Delta T=0.50K$ (red) and  (b) vs. $\Delta T$ with $Z \to 0$ (blue), $Z=1.0$ (red), $Z=5.0$ (orange) and at zero bias voltage ($V_1=V_2=0$) and $\Bar{T} = 3.0K$. We consider the superconducting gap $\Delta_0 = 1.76 k_B T_c$, where $T_c = 18K$.}
\label{fig100}
\end{figure}

\begin{figure}[h!]
\includegraphics[scale=.37]{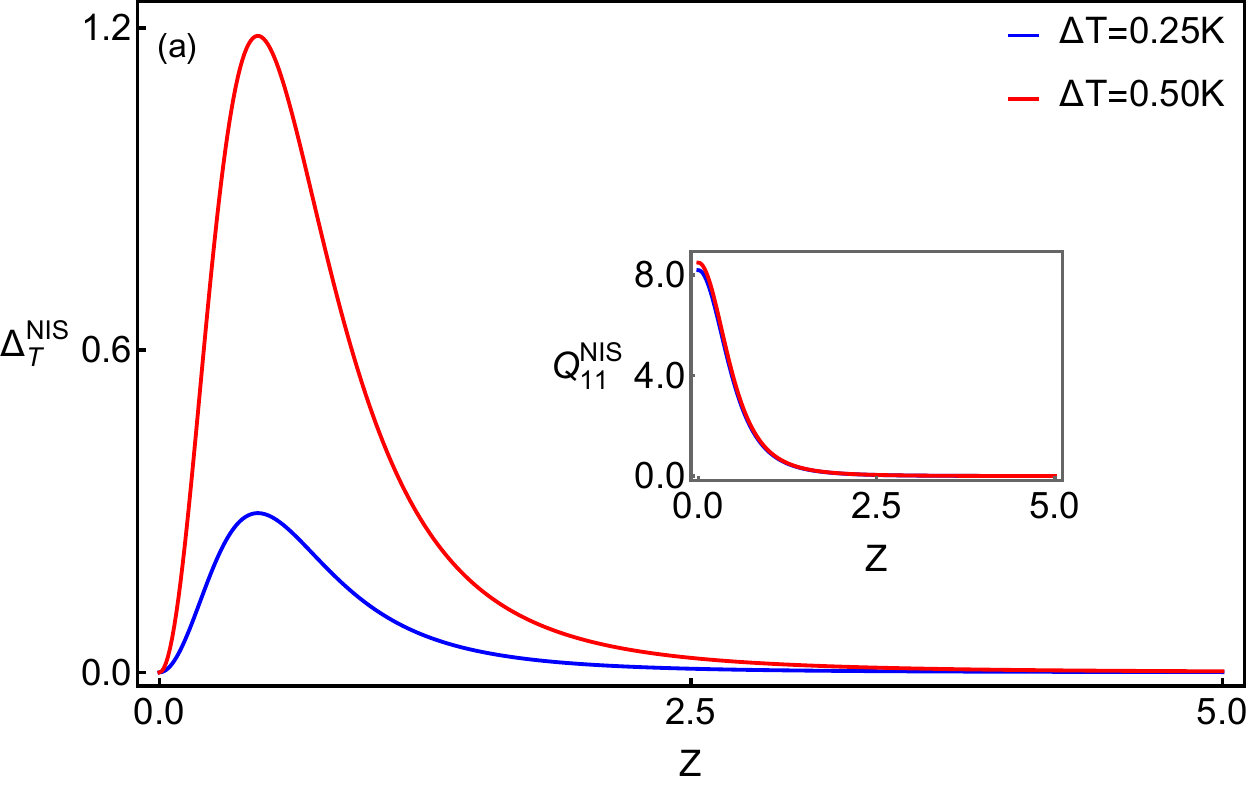}
\includegraphics[scale=.37]{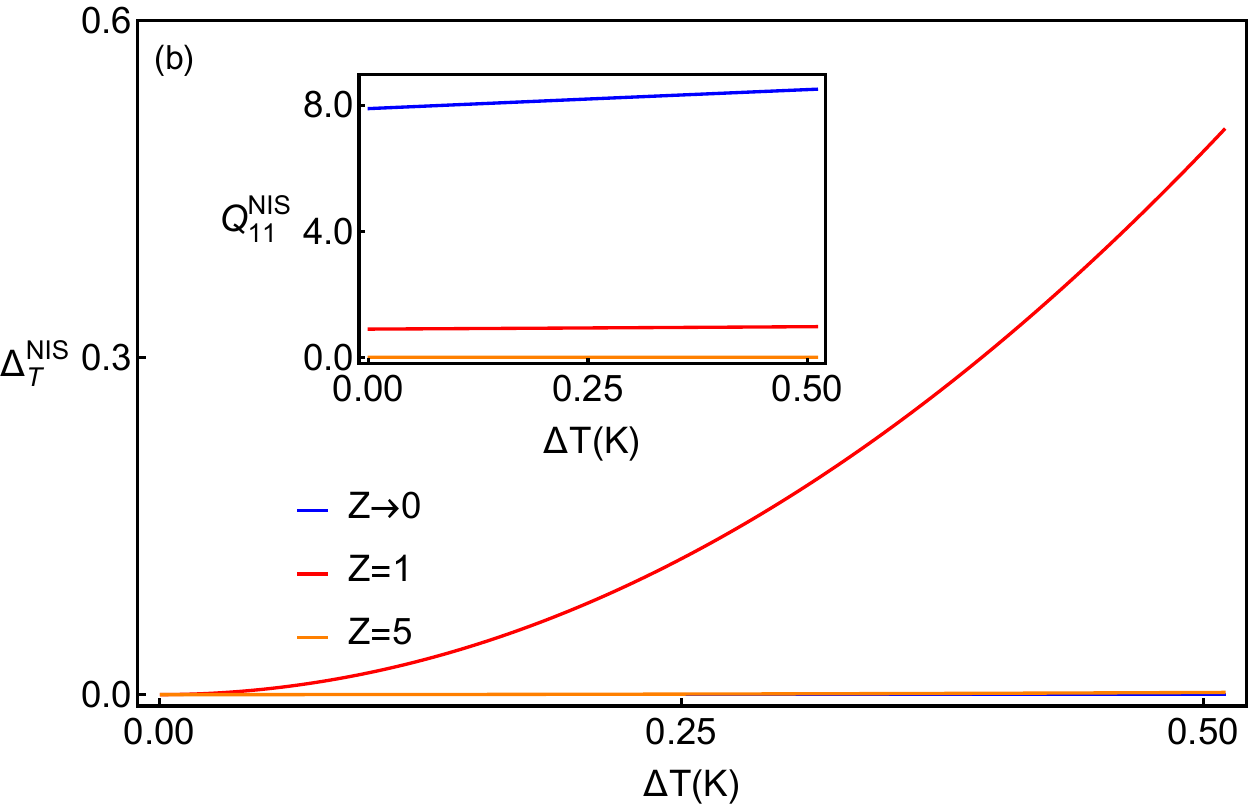}
\caption{$\Delta_T^{NIS}$ noise (Quantum  noise $Q^{NIS}_{11}$ in units of $\frac{2e^2}{h} k_B \bar{T}$ in inset) in units of $\frac{2e^2}{h} 10^{-2} k_B \Bar{T}$ (a)  vs. $Z$, with $\Delta T=0.25K$ (blue), $\Delta T=0.50K$ (red) and  (b) vs. $\Delta T$ with $Z \to 0$ (blue), $Z=1.0$ (red), $Z=5.0$ (orange) and at zero bias voltage ($V_1=V_2=0$) and $\Bar{T} = 3.0K$. We consider the superconducting gap $\Delta_0 = 1.76 k_B T_c$, where $T_c = 18K$.}
\label{fig101}
\end{figure}

\end{widetext}

{In Fig.~\ref{fig101}, we present the total quantum noise and $\Delta_{T}^{NIS}$, as a function of both $Z$ and $\Delta T$. In Fig.~\ref{fig101}(a), $\Delta_T^{NIS}$ varies non-monotonically with $Z$, demonstrating a more complex dependence on the barrier strength. In contrast, the behavior of $Q_{11}^{NIS}$, as shown in the inset of Fig.~\ref{fig101}(a) exhibits a {distinct nature to that of} $\Delta_T^{NIS}$, see Fig.~\ref{fig101}(a). $Q_{11}^{NIS}$ attains its maximum in the transparent limit ({$Z \to 0$}). As $Z$ increases, $Q_{11}^{NIS}$ gradually decreases, eventually {vanishing} in the tunneling limit. This trend highlights the suppression of quantum noise as the barrier strength $Z$ increases. }

{Additionally, we explore the behavior of $\Delta_{T}^{NIS}$ as a function of $\Delta T$ for specific values of $Z$, namely 0, 1, and 5, see Fig. \ref{fig101}(b). It {is quite small} at {$Z \to 0$} but increases quadratically with $\Delta T$ for $Z = 1$ and $Z = 5$. Conversely, $Q_{11}^{NIS}$ behaves quite differently, see inset of Fig. \ref{fig101}(b). At {$Z \to 0$}, $Q_{11}^{NIS}$ is the {largest}, and it progressively reduces for $Z = 1$ and $Z = 5$. Interestingly, $Q_{11}^{NIS}$ {increases linearly} with respect to $\Delta T$ for all $Z$ values {in contrast to the NIN case}. These observations provide a comprehensive understanding of the distinct behaviors of quantum noise and $\Delta_T$ noise in NIS junctions, shedding light on the interplay between barrier strength, thermal gradients, and noise characteristics. The ratio \(\frac{Q_{11}^{NIS}}{Q_{11}^{NIN}}\) {remains greater than 2, indicating a strong influence of Andreev reflection on the noise characteristics}.
}

\begin{widetext}

\begin{figure}[h!]
\includegraphics[scale=.32]{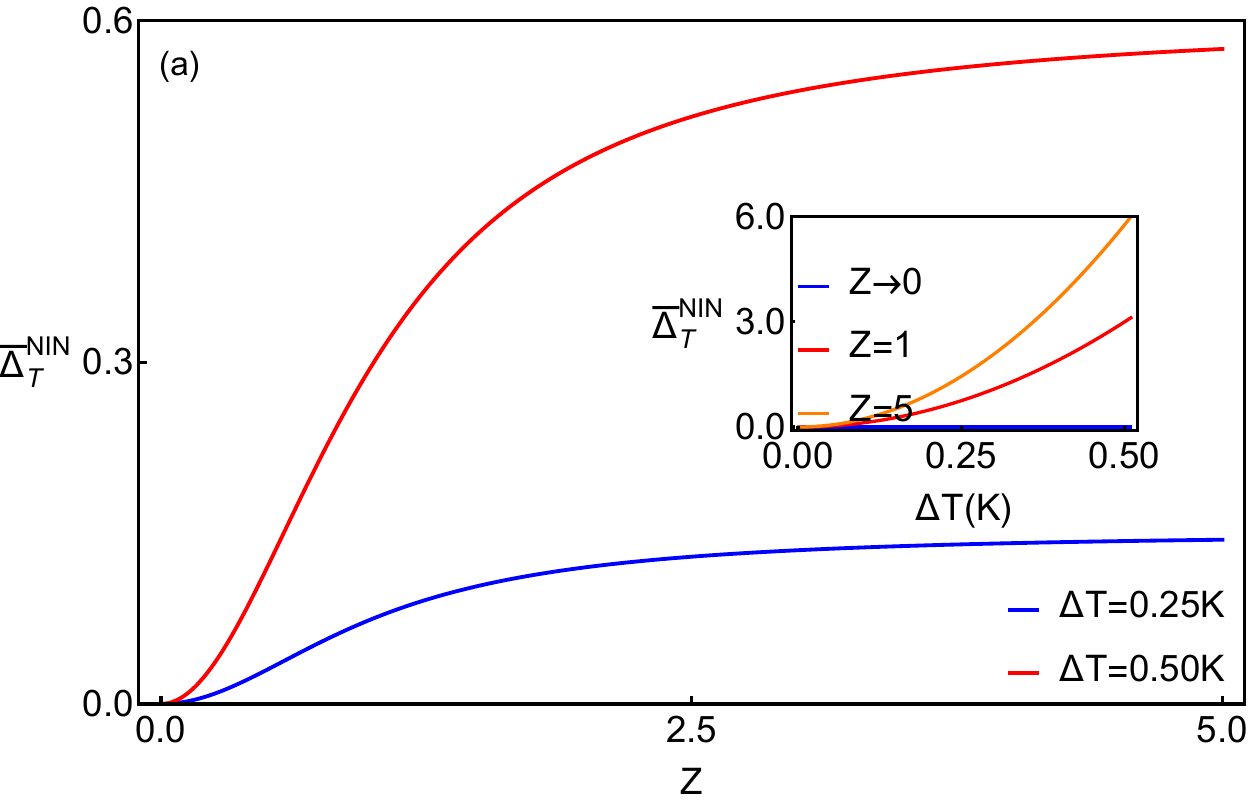}
\includegraphics[scale=.33]{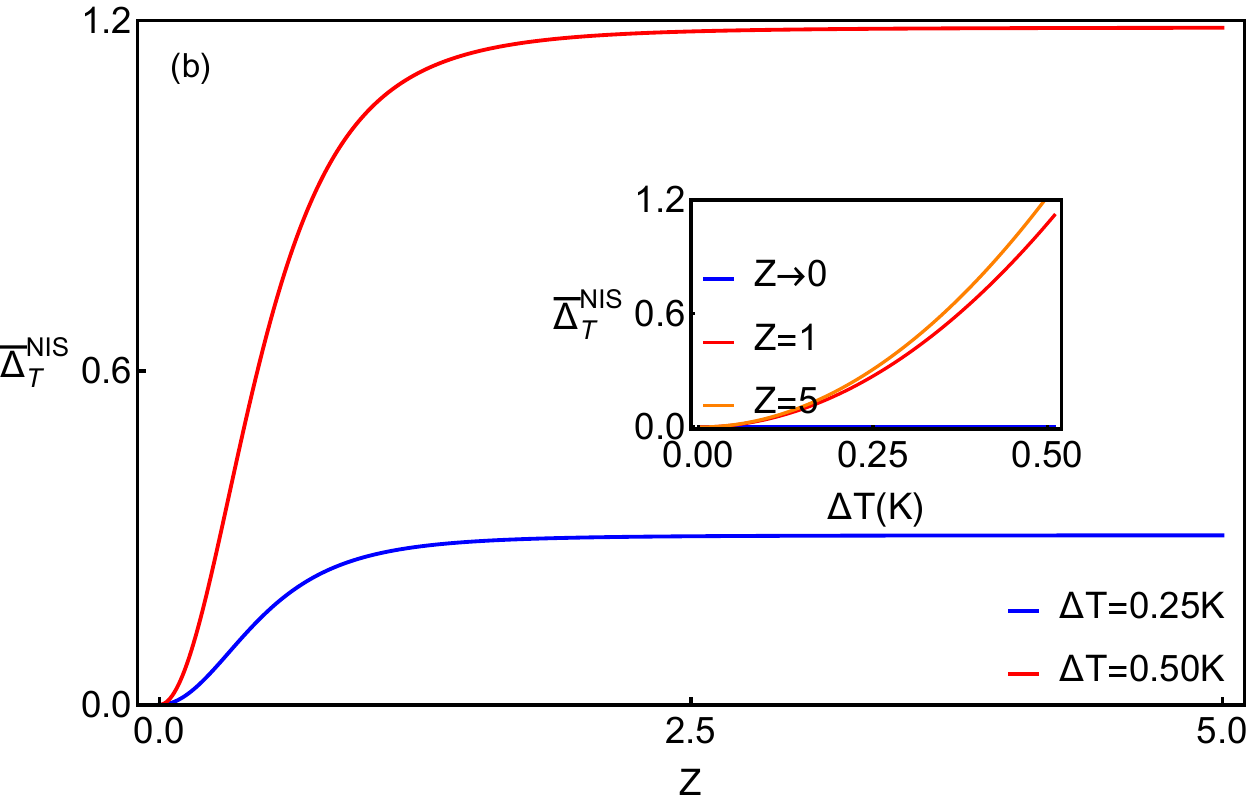}
\caption{{(a) $\bar{\Delta}_T^{NIN}$, (b) $\bar{\Delta}_T^{NIS}$} in units of $10^{-2} k_B \Bar{T}$ vs. $Z$ with $\Delta T=0.25K$ (blue), $\Delta T=0.50K$ (red) and vs. $\Delta T$ {(in insets)} with {$Z \to 0$} (blue), $Z=1.0$ (red), $Z=5.0$ (orange) and at zero bias voltage ($V_1=V_2=0$) and $\bar{T} = 3.0K$. We consider the superconducting gap $\Delta_0 = 1.76 k_B T_c$, where $T_c = 18K$.}
\label{fig102}
\end{figure}

\end{widetext}

{We also plot the normalized $\Delta_T$ noise, $\bar{\Delta}_T^{NIN} = \frac{\Delta_T^{NIN}}{G_{NIN}}$, as function of the barrier strength $Z$ for specific temperature bias at $\Delta T = \left(0.25K, 0.50K; \text{with $\Delta T \ll \bar{T} = 3.0K$}\right)$ for an NIN junction, as shown in Fig. \ref{fig102}. Here, $G_{NIN}$ denotes the charge conductance, whose expression is provided below Eq. (\ref{A_I}) in Appendix \ref{App_I}. Our analysis reveals that $\bar{\Delta}_T^{NIN}$ increases monotonically with $Z$ for temperature biases $\Delta T = 0.25K$ and $0.50K$, as illustrated in Fig. \ref{fig103}(a). Furthermore, as depicted in inset of Fig. \ref{fig102}(a), $\bar{\Delta}_T^{NIN}$ exhibits a quadratic dependence on $\Delta T$ for $Z = 1$ and $5$, whereas it is quite small for {$Z \to 0$}.}  

{Similarly, for the NIS junction, we find that the normalized $\Delta_T$ noise, defined as $\bar{\Delta}_T^{NIS} = \frac{\Delta_T^{NIS}}{G_{NIS}}$ is very small in the transparent limit ($Z \to 0$), as shown in Fig. \ref{fig102}(b). As $Z$ increases, $\bar{\Delta}_T^{NIS}$ also exhibits a monotonic behavior just like NIN case. Notably, in the intermediate limit ($Z = 1$) as well as tunnel limit ($Z = 5$), $\bar{\Delta}_T^{NIS}$ is more than twice the value of $\bar{\Delta}_T^{NIN}$, as illustrated in Fig. \ref{fig102}(b). This enhancement arise due to Andreev reflection, making $\bar{\Delta}_T^{NIS}$ a useful probe for detecting Andreev reflection in NIS junctions. Here, $G_{NIS}$ represents the conductance in a NIS junction, whose expression is provided below Eq. (\ref{A_I}) in Appendix \ref{App_I}. Inset of Figure \ref{fig102}(b) presents the variation of $\bar{\Delta}_T^{NIS}$ with respect to $\Delta T$ for different values of $Z$. We observe that, akin to the NIN junction, $\bar{\Delta}_T^{NIS}$ follows a quadratic dependence on $\Delta T$, though its magnitude remains almost twice that of the NIN junction.}

\section{Analysis}
\label{analysis}

{
% In this section, we analyze the ratio of $\Delta_T$ noise $\left(\frac{\Delta_T^{NIS}}{\Delta_T^{NIN}}\right)$ and normalized $\Delta_T$ noise $\left(\frac{\bar{\Delta}_T^{NIS}}{\bar{\Delta}_T^{NIN}}\right)$ both in NIS to that in a NIN junction.

{In this section, we analyze the ratio of $\Delta_T$ noise $\left(\frac{\Delta_T^{NIS}}{\Delta_T^{NIN}}\right)$ and normalized $\Delta_T$ noise $\left(\frac{\bar{\Delta}_T^{NIS}}{\bar{\Delta}_T^{NIN}}\right)$ in NIS to that in a NIN junction for different barrier strengths ($z\to0,1,5$) and temperature biases ($\Delta T = 0.25K, 0.5K$).}

}

{In this work, we define the temperature of the normal metal as $T_1 = \Bar{T} + \frac{\Delta T}{2}$ and the temperature of the superconductor as $T_2 = \Bar{T} - \frac{\Delta T}{2}$, where $\Bar{T}$ represents the average temperature of the system, and $\Delta T$ denotes the temperature bias. We specifically consider the regime where $\Delta T \ll \Bar{T}$. We see that the charge quantum noise {in NIN junction is} almost constant with temperature bias, see inset of Fig. \ref{fig100}(b){, whereas the charge quantum noise in NIS junction is more than 2 than that of NIN junction, see inset of Fig. \ref{fig101}(b).} However, the $\Delta_T$ noise (both $\Delta_T^{NIN}$ and $\Delta_T^{NIS}$) and normalized $\Delta_T$ noise (both $\bar{\Delta}_T^{NIN}$ and $\bar{\Delta}_T^{NIS}$) changes quadratically with temperature bias $\Delta T$. } 

\begin{comment}
\begin{widetext}

\begin{table}[h]
\caption{Behavior of quantum noise $Q_{11}^{NIN}$, $Q_{11}^{NIS}$, $\Delta_T$ noise $\Delta_{T}^{NIN}$, $\Delta_{T}^{NIS}$ and normalized $\Delta_T$ noise $\bar{\Delta}_{T}^{NIN}$, $\bar{\Delta}_{T}^{NIS}$ as a function of temperature bias ($\Delta T$) in the transparent ($Z \to 0$), intermediate ($Z = 1$) and tunneling limit ($Z = 5$).}
\centering
\scalebox{1.50}{
\renewcommand{\arraystretch}{1.5}
\begin{tabular}{|c|c|c|c|c|c|c|c|}
\hline
 & $Q_{11}^{NIN}$ & $Q_{11}^{NIS}$ & $\Delta_{T}^{NIN}$ & $\Delta_{T}^{NIS}$ & $\bar{\Delta}_{T}^{NIN}$ & $\bar{\Delta}_{T}^{NIS}$ \\ 
\hline
$Z \to 0$ & Constant & Constant & Quadratic & Quadratic & Quadratic & Quadratic \\  \hline
$Z = 1$   & Constant & Constant & Quadratic & Quadratic & Quadratic & Quadratic \\  \hline
$Z = 5$   & Constant & Constant & Quadratic & Quadratic & Quadratic & Quadratic \\  
\hline
\end{tabular}}
\label{Table1}
\end{table}
\end{widetext}

\end{comment}

\begin{widetext}

\begin{figure}[h!]
\includegraphics[scale=.31]{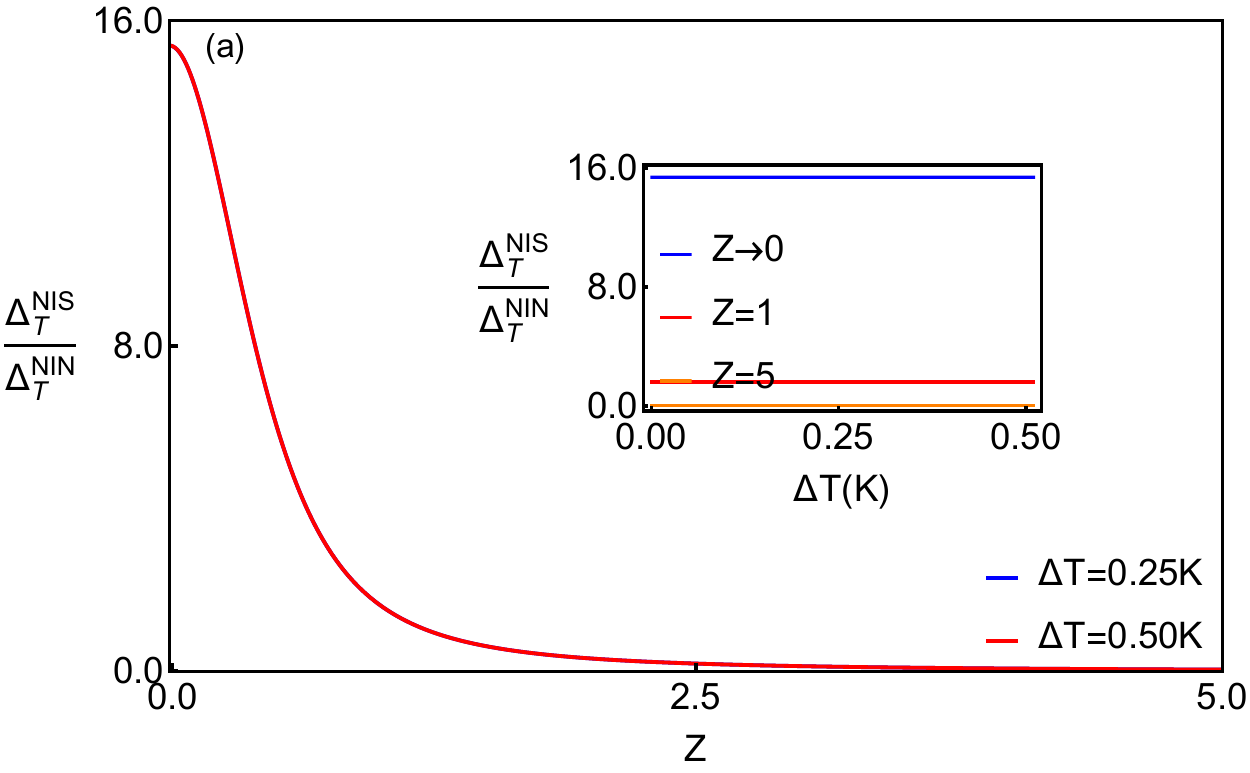}
\includegraphics[scale=.3]{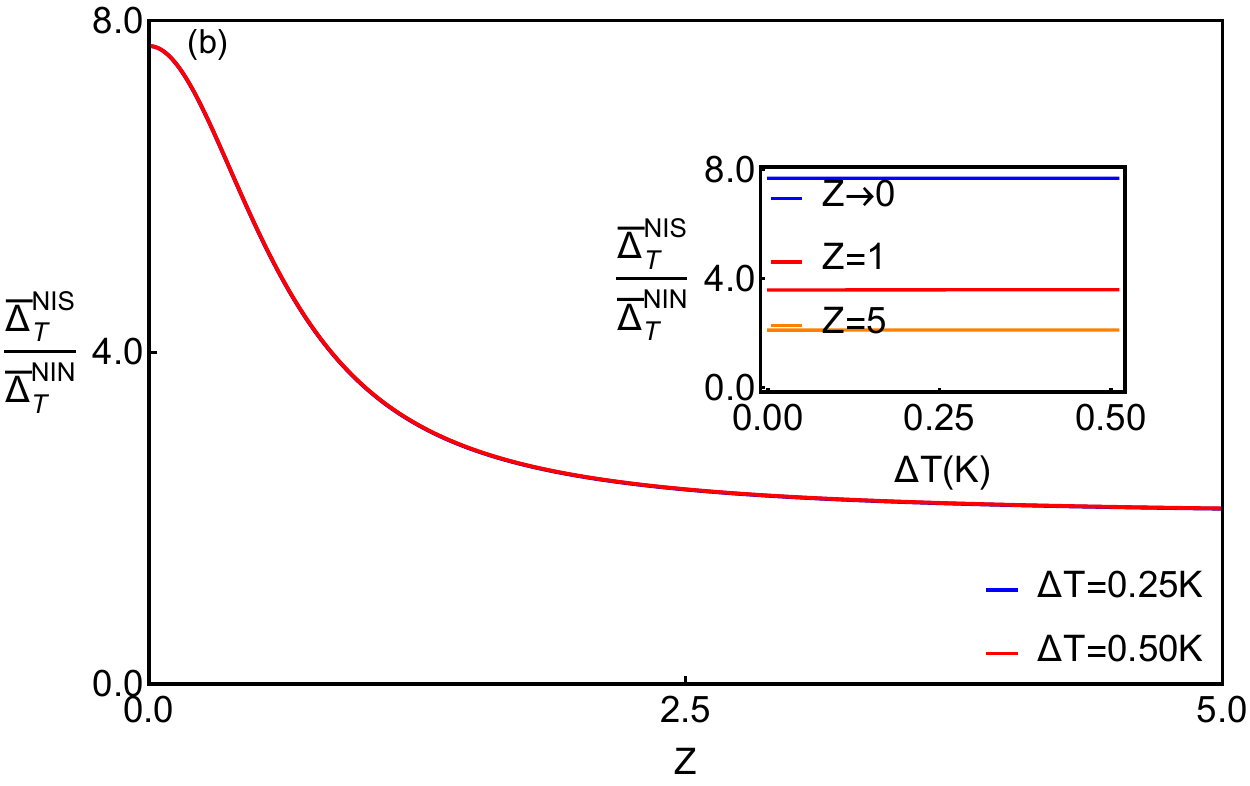}
\caption{{(a) $\frac{\Delta_T^{NIS}}{\Delta_T^{NIN}}$, (b) $\frac{\bar{\Delta}_T^{NIS}}{\bar{\Delta}_T^{NIN}}$} vs. $Z$ with $\Delta T=0.25K$ (blue), $\Delta T=0.50K$ (red) and vs. $\Delta T$ (in insets) with $Z \to 0$ (blue), $Z=1.0$ (red), $Z=5.0$ (orange) and at zero bias voltage ($V_1=V_2=0$) and $\bar{T} = 3.0K$. We consider the superconducting gap $\Delta_0 = 1.76 k_B T_c$, where $T_c = 18K$.}
\label{fig103}
\end{figure}
    
\begin{table}[h!]
\centering
\renewcommand{\arraystretch}{2} % Increase row height
\setlength{\tabcolsep}{10pt} % Increase column spacing
\caption{Ratio of $\Delta_T$ noise $\frac{\Delta_{T}^{NIS}}{ \Delta_{T}^{NIN}}$ and normalized $\Delta_T$ noise $\frac{\bar{\Delta}_{T}^{NIS}}{\bar{\Delta}_{T}^{NIN}}$ for $Z \to 0$ (transparent), 1 (intermediate), 5 (tunneling).}
\scalebox{1.20}{
\begin{tabular}{|c|c|c|c|c|c|}
\hline
\multicolumn{2}{|c|}{$Z \to 0$} 
& \multicolumn{2}{c|}{$Z = 1$} 
& \multicolumn{2}{c|}{$Z$ = 5} 
\\ \hline
$\frac{\Delta_{T}^{NIS}}{\Delta_{T}^{NIN}}$ & $\frac{\bar{\Delta}_{T}^{NIS}}{\bar{\Delta}_{T}^{NIN}}$
& $\frac{\Delta_{T}^{NIS}}{\Delta_{T}^{NIN}}$ & $\frac{\bar{\Delta}_{T}^{NIS}}{\bar{\Delta}_{T}^{NIN}}$
& $\frac{\Delta_{T}^{NIS}}{\Delta_{T}^{NIN}}$ & $\frac{\bar{\Delta}_{T}^{NIS}}{\bar{\Delta}_{T}^{NIN}}$
\\ \hline
$\lesssim 16$ & $\lesssim 8$ & $\lesssim$ 2 & $\lesssim$ 4 & $\simeq$ 0 & $\lesssim2$ \\ \hline
\end{tabular}}
\label{Table2}
\end{table}

\end{widetext}

{In Fig.~\ref{fig103}(a), we plot the ratio $\frac{\Delta_T^{NIS}}{\Delta_T^{NIN}}$, as a function of the barrier strength \( Z \) for two distinct temperature biases, \(\Delta T = 0.25\,\text{K}\) and \(0.5\,\text{K}\). For almost transparent junction, specifically \( Z \to 0 \), this ratio reaches a significantly {large} value of around 16, see Table \ref{Table2}. This behavior highlights the pronounced influence of Andreev reflection for transport junctions, where the NIS junction exhibits much greater $\Delta_T$ noise compared to NIN junction. As \( Z \) increases, the barrier suppresses Andreev reflection, and the ratio gradually decreases. We provide the value of $\frac{\Delta_T^{NIS}}{\Delta_T^{NIN}}$ in different limits such as intermediate limit ($Z = 1$) and tunneling limit ($Z = 5$) in Table \ref{Table2}. For sufficiently large \( Z \), i.e., tunnel barrier limit, the ratio saturates to zero, indicating that the $\Delta_T$ noise in the NIS junction is negligible relative to that in the NIN junction, see Fig. \ref{fig103}(a).}

{Similarly, in {inset of} Fig.~\ref{fig103}(a), we analyze, $\frac{\Delta_T^{NIS}}{\Delta_T^{NIN}}$, but now as a function of the temperature bias \(\Delta T\) for three fixed barrier strengths \( Z \): {\( Z \to 0 \)}, \( Z = 1 \), and \( Z = 5 \). The results reveal that the ratio remains almost constant with varying \(\Delta T\), implying that the temperature bias has a minimal direct impact on the relative $\Delta_T$ noise between the two junction types. However, the magnitude of the ratio depends strongly on \( Z \). At {\( Z \to 0 \)}, {for transparent junction}, the ratio is significantly large, around 16. {The analytical derivation of the ratio of $\Delta_T$ noise at $Z \to 0$ is given in Appendix \ref{ratio}.} For \( Z = 1 \), {intermediate barrier strength}, the ratio drops below {2} due to the suppression of Andreev reflection by the barrier. In the tunnel barrier limit (\( Z = 5 \)), the suppression is even more pronounced, and the ratio decreases further, approaching zero. }

{We further analyze our results by examining the ratio of normalized $\Delta_T$ noise, $\frac{\bar{\Delta}_T^{NIS}}{\bar{\Delta}_T^{NIN}}$, as a function of the barrier strength $Z$ and temperature bias $\Delta T$. Our findings highlight the significant impact of Andreev reflection on charge transport. In the transparent limit ($Z \to 0$), the ratio $\frac{\bar{\Delta}_T^{NIS}}{\bar{\Delta}_T^{NIN}}$ is approximately 8, as shown in Table \ref{Table2}. As $Z$ increases, the ratio decreases for both temperature biases $\Delta T = 0.25K$ and $0.50K$, as illustrated in Fig. \ref{fig103}(b). The specific values of $\frac{\bar{\Delta}_T^{NIS}}{\bar{\Delta}_T^{NIN}}$ in the intermediate limit ($Z = 1$) and the tunneling regime ($Z = 5$) are provided in Table \ref{Table2}. In the transparent limit \(\left(Z \to 0\right)\), the ratio \(\frac{\Delta_T^{NIS}}{\Delta_T^{NIN}}\) is twice that of \(\frac{\bar{\Delta}_T^{NIS}}{\bar{\Delta}_T^{NIN}}\) (see Fig. \ref{fig103}). This can be understood as follows: The ratio of the normalized noise, \(\frac{\bar{\Delta}_T^{NIS}}{\bar{\Delta}_T^{NIN}}\), is given by  
$\frac{\bar{\Delta}_T^{NIS}}{\bar{\Delta}_T^{NIN}} = \frac{G_{NIN}}{G_{NIS}} \frac{\Delta_T^{NIS}}{\Delta_T^{NIN}}$.
In the transparent limit (\(Z \to 0\)), where \(\frac{G_{NIN}}{G_{NIS}} \to \frac{1}{2}\) as shown in \cite{BTK}, this expression simplifies to approximately  
$\frac{\bar{\Delta}_T^{NIS}}{\bar{\Delta}_T^{NIN}} \approx \frac{1}{2} \frac{\Delta_T^{NIS}}{\Delta_T^{NIN}}.
$
Consequently,  
$\frac{\Delta_T^{NIS}}{\Delta_T^{NIN}} \approx 2 \frac{\bar{\Delta}_T^{NIS}}{\bar{\Delta}_T^{NIN}}.$
 However, in tunneling limit, $Z \to $ large, i.e., $Z \to 5$, \(\frac{\Delta_T^{NIS}}{\Delta_T^{NIN}}\) approaches zero, while $\bar{\Delta}_T^{NIS}$ remains finite and is approximately twice of $\bar{\Delta}_T^{NIN}$.}

{Additionally, we investigate the dependence of ratio of normalized $\Delta_T$ noise, i.e., $\frac{\bar{\Delta}_T^{NIS}}{\bar{\Delta}_T^{NIN}}$ on $\Delta T$ for $Z \to 0$, $Z = 1$, and $Z = 5$, as shown in the inset of Fig. \ref{fig103}(b). Our analysis reveals that the ratio remains nearly constant with respect to $\Delta T$ across all considered values of $Z$. Specifically, at $Z \to 0$, $\frac{\bar{\Delta}_T^{NIS}}{\bar{\Delta}_T^{NIN}}$ is approximately 8, while for $Z = 1$ and $Z = 5$, it decreases from this initial value, with their values listed in Table \ref{Table2}. Normalized $\Delta_T$ noise is akin to the Fano factor and is a measure of charge transport. Therefore, in the transparent limit ($Z \to 0$), the effective charge transported is around $8e$. In the intermediate regime, it is around $4e$, while in the tunneling limit ($Z \to 5$), it remains approximately $2e$ (see Fig. \ref{fig103}(b) and Table \ref{Table2}). These findings emphasize the crucial role of barrier strength in shaping both the absolute and normalized $\Delta_T$ noise in NIS and NIN junctions, offering new insights into the interplay between quantum transport and thermal fluctuations.}

{In a recent study conducted in Ref.~\cite{DeltaNS}, the focus is on calculating the total quantum noise, which erroneously identifies as $\Delta_T$ noise. This is a factual inaccuracy. In reality, $\Delta_T$ noise is a distinct component of quantum noise, specifically the shot noise-like contribution that arises due to a temperature gradient at zero charge current transported, see Ref. \cite{atomicscaleexpt} for the correct definition. Shot noise-like contributions are of significant importance in mesoscopic systems, and distinguishing them from the total quantum noise is crucial for a deeper understanding of noise mechanisms.}

{Ref.~\cite{DeltaNS} reported that the ratio of quantum noise— which incorrectly identifies as $\Delta_T$ noise— in a normal metal-insulator-superconductor (NIS) junction is twice that in a normal metal-insulator-normal metal (NIN) junction in the transparent regime. While their numerical result is correct, their analysis fails to distinguish the essential contribution of shot noise, which is defined as the $\Delta_T$ noise in absence of charge current. This omission is significant because, in a system with a temperature bias, the total quantum noise is dominated by thermal noise, rendering shot noise (or, $\Delta_T$ noise) considerably smaller—in our case, by a factor of 100 in absence of charge current. Consequently, without isolating the shot noise component, their study essentially reproduces information already contained in the conductance, since thermal noise follows the same behavior as conductance in such setups. In other words, the work of Ref. \cite{DeltaNS} does not provide new insights beyond what can be inferred from standard conductance measurements, as it does not properly address $\Delta_T$ noise. In contrast, our work presents the first rigorous and systematic analysis of $\Delta_T$ noise, carefully distinguishing it from the total quantum noise and highlighting its distinct physical implications. By thoroughly examining the impact of temperature bias and Andreev reflection on $\Delta_T$ noise, our study introduces a fundamentally new perspective that was completely absent in the work of Ref. \cite{DeltaNS}. }

% {It is also important to note that Ref.~\cite{DeltaNS} considers the regime $\Delta T > \Bar{T}$, which is nonlinear and uses the methodology of linear response to calculate noise. In nonlinear regime, one can not use linear response methods. We assume $\Delta T \ll \Bar{T}$ as we are using linear response methods. Our work correctly defines and evaluates $\Delta_T$ noise, which is not addressed in Ref.~\cite{DeltaNS}.} 

{
It is also important to note that Ref. \cite{DeltaNS} considers the regime $\Delta T > \Bar{T}$, which is nonlinear but uses linear response techniques to calculate quantum noise. The analysis of quantum noise in the nonlinear response regime is not as straightforward as the linear response regime. In our manuscript, we assume $\Delta T \ll \Bar{T}$ as we are in linear response regime and analyze our setup using the linear response Landaeur-Buttiker formalism. Our manuscript defines and evaluates $\Delta_T$ noise, which is not addressed in Ref. \cite{DeltaNS}. More details on applying nonlinear response in mesoscopic conductors can be found in Refs. \cite{Buttiker1993} and \cite{Lopez2013}.
}

{Our paper establishes a significant and previously unreported result: in the transparent limit, the $\Delta_T$ noise in an NIS junction is around 16 times larger than that in an NIN junction. Furthermore, we demonstrate that the normalized $\Delta_T$ noise in the NIS junction is more than times greater than in the NIN junction in a transparent junction. These quantitative findings are of fundamental importance and have never been reported in the literature before.}

\subsection{{Quantum thermal noise vs. $\Delta_T$ noise}}
\label{analysisA}

\begin{widetext}

\begin{figure}[h!]
\includegraphics[scale=.32]{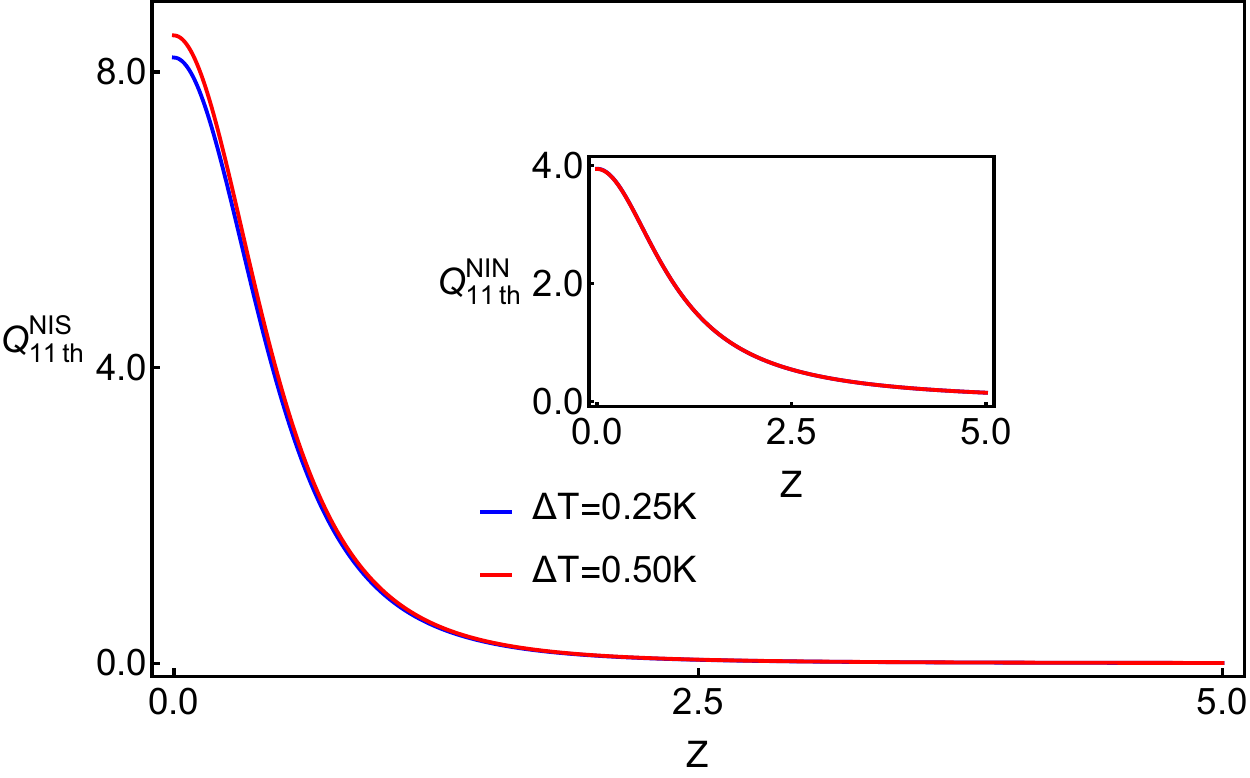}
\includegraphics[scale=.32]{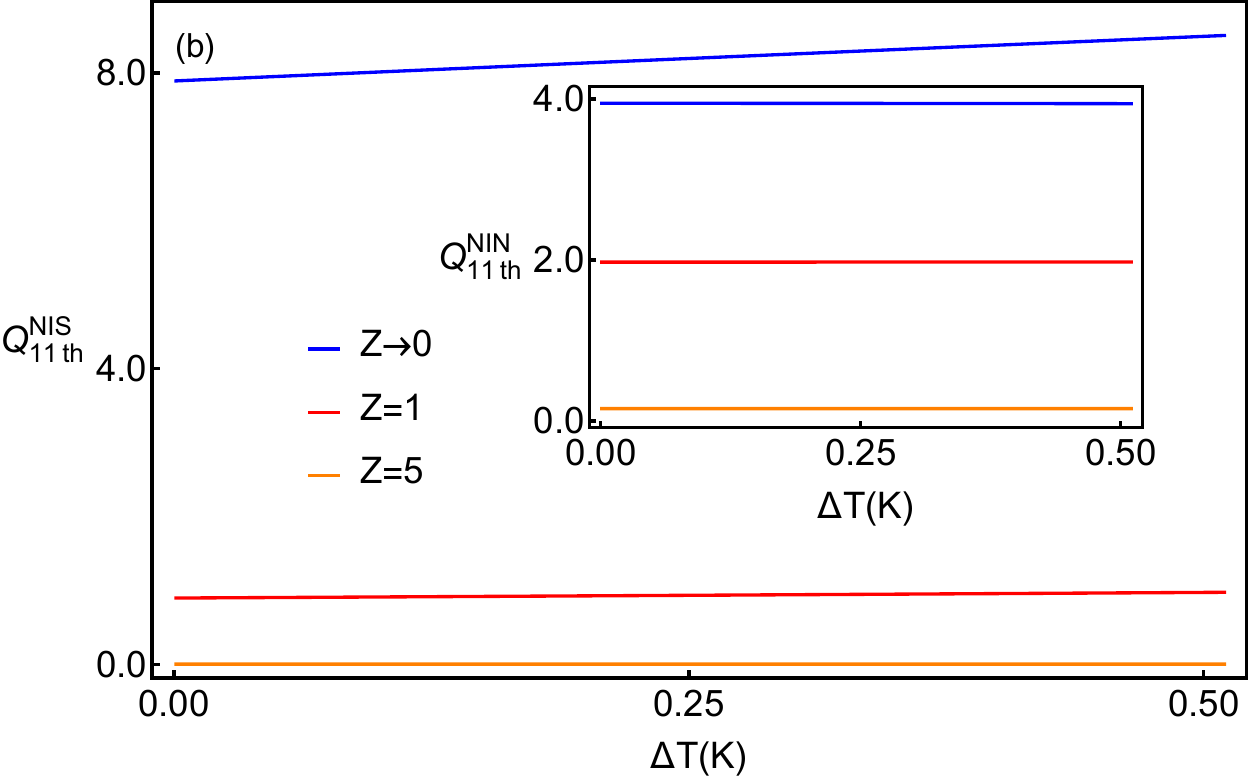}
\caption{Quantum thermal noise in NIS ($Q^{NIS}_{11th}$) (Quantum thermal noise in NIN ($Q^{NIN}_{11th}$) in inset) in units of $\frac{2e^2}{h} k_B \bar{T}$ (a)  vs. $Z$ with $\Delta T=0.25K$ (blue), $\Delta T=0.50K$ (red) and (b) vs. $\Delta T$ with $Z \to 0$ (blue), $Z=1.0$ (red), $Z=5.0$ (orange) and at zero bias voltage ($V_1=V_2=0$). We consider the superconducting gap $\Delta_0 = 1.76 k_B T_c$, where $T_c = 18K$.}
\label{fig104}
\end{figure}
\end{widetext}

{As discussed in Ref. \cite{generalbound}, the total charge quantum noise at zero voltage bias and finite temperature bias, with zero average charge current, consists of two primary components: a shot noise-like contribution, denoted as $Q_{11sh}$, and a thermal noise contribution, denoted as $Q_{11th}$. It has been shown that, under such conditions, the shot noise-like contribution $Q_{11sh}$ is much smaller than the thermal noise contribution $Q_{11th}$. This specific component, $Q_{11sh}$, is also referred to as the charge $\Delta_T$ noise in the literature \cite{generalbound, atomicscaleexpt}. Consequently, the total quantum noise is predominantly determined by the thermal noise, meaning that we can approximate it as $Q_{11} \approx Q_{11th}$ as shown in Fig. \ref{fig104}.}

{As shown in Eq. (\ref{eqn:S_NIN}), the thermal noise-like contribution in NIN junction is}

{
\begin{equation}
\begin{split}
Q_{11th}^{NIN} &= \frac{4e^2}{h} \int_{-\infty}^{\infty} \mathcal{T}_{NIN} \{F1e + F2e \}dE
\end{split}
\end{equation}
}

{where $Fie = f_{ie} (1-f_{ie})$ for $i \in \{1,2 \}$.  Now, using $f_{1e} (1 - f_{1e}) = k_B T_1 \left(\frac{-\partial f_{1e}}{\partial E}\right)$ and $f_{2e} (1 - f_{2e}) = k_B T_2 \left(\frac{-\partial f_{2e}}{\partial E}\right)$, one can Eq. (14) to be}

{
\begin{equation}
\begin{split}
    Q_{11th}^{NIN} &= \frac{4e^2}{h} \mathcal{T}_{NIN} \int_{-\infty}^{\infty} dE \bigg(k_B \left(\bar{T} + \frac{\Delta T}{2}\right) \left(\frac{-\partial f_{1e}}{\partial E}\right) \\ & + k_B \left(\bar{T} - \frac{\Delta T}{2}\right) \left(\frac{-\partial f_{2e}}{\partial E}\right) \bigg)\\
    &= \frac{8e^2}{h} \mathcal{T}_{NIN} k_B \bar{T},
\end{split}    
\end{equation}
}

{$\mathcal{T}_{NIN}$ is independent of energy $E$ and thus is taken out of the integral and we utilize the property $\int_{-\infty}^{\infty} dE \left(\frac{-\partial f_{1e(2e)} }{\partial E}\right) = 1$. We observe from Eq. (15) that $Q_{11th}^{NIN}$ is proportional to only the average temperature $\bar{T}$ and not on temperature bias $\Delta T$. When $Z \to 0$, $\mathcal{T}_{NIN} \to 1$, we get $Q_{11th}^{NIN} (Z \to 0)$ = $\frac{8e^2}{h} k_B \bar{T}$.}

{In the similar fashion, one can also analyze why the thermal noise-like contribution in NIS junction, i.e., $Q_{11th}^{NIS}$ varies with $\Delta T$. According to Eq. (\ref{eqn:S_NIS}), $Q_{11th}^{NIS}$ is given as}
\begin{widetext}

{
\begin{equation}
    \begin{split}
        Q_{11th}^{NIS} &= \frac{4e^2}{h} \int_{-\infty}^{\infty} dE \bigg((1 - \mathcal{B} +  3 \mathcal{A}) F1e +  (1 - \mathcal{B} - \mathcal{A}) F2e\bigg)\\
        &= \frac{4e^2}{h} \int_{-\infty}^{\infty} dE \bigg((1 - \mathcal{B} + \mathcal{A}) (F1e + F2e)\bigg) + \frac{4e^2}{h} \int_{-\infty}^{\infty} dE \, \, \, \, \, \, 2 \mathcal{A} (F1e - F2e)\\
        &= \frac{4e^2}{h} \int_{-\infty}^{\infty} dE \bigg((1 - \mathcal{B} + \mathcal{A}) \bigg(k_B \left(\bar{T} + \frac{\Delta T}{2}\right) \left(\frac{-\partial f_{1e}}{\partial E}\right) + k_B \left(\bar{T} - \frac{\Delta T}{2}\right) \left(\frac{-\partial f_{2e}}{\partial E}\right) \bigg) \bigg)\\& + \frac{4e^2}{h} \int_{-\infty}^{\infty} dE \,\,\,\,\, 2 \mathcal{A} \bigg(k_B \left(\bar{T} + \frac{\Delta T}{2}\right) \left(\frac{-\partial f_{1e}}{\partial E}\right) - k_B \left(\bar{T} - \frac{\Delta T}{2}\right) \left(\frac{-\partial f_{2e}}{\partial E}\right) \bigg)
    \end{split}
\end{equation}
}
\end{widetext}
{For the NIS junction, the scattering probabilities $\mathcal{A}$, $\mathcal{B}$ are dependent on $E$, therefore cannot be taken out of the integral. However, for $Z \to 0$, $\mathcal{A} \to 1$ and $\mathcal{B} \to 0$ and in this case, one can take $\mathcal{A}$ and $\mathcal{B}$ out of the inegral. Therefore, Eq. (16) is reduced to}
{
        $Q_{11th}^{NIS} (Z \to 0) = \frac{8e^2}{h} k_B (2 \bar{T} + \Delta T)$
.}
{We observe that $Q_{11th}^{NIS} (Z \to 0)$ is linear with $\Delta T$ unlike $Q_{11th}^{NIN} (Z \to 0)$. Now, the ratio $\frac{Q_{11th}^{NIS} (Z \to 0)}{Q_{11th}^{NIN} (Z \to 0)}$ is $\left(2 + \frac{\Delta T}{\bar{T}}\right)$, for linear response as $\Delta T \ll \bar{T}$, therefore $\frac{Q_{11th}^{NIS}}{Q_{11th}^{NIN}} \simeq 2$ in the transparent limit.}

{$\Delta_T$ noise quadratically changes with temperature bias $\Delta T$ (see, Figs. \ref{fig100} and \ref{fig101}), but is 100 times weaker than thermal noise. We have plotted the specific $\Delta_T$ noise contribution in Figs. \ref{fig100} and \ref{fig101}, which shows the quadratic dependence of $\Delta T$. {At $Z \to 0$, $\Delta_T^{NIN}$ and $\Delta_T^{NIS}$ almost vanishes, but their ratio $\frac{\Delta_T^{NIS}}{\Delta_T^{NIN}} \lesssim 16$. As $Z$ increases, $\frac{\Delta_T^{NIS}}{\Delta_T^{NIN}}$ reduces, see Fig. \ref{fig103}(a) and Table \ref{Table2}, as in the intermediate limit ($Z = 1$), it is around 2 and in the tunneling limit ($Z = 5$), it vanishes.}}
{We also examine the normalized $\Delta_T$ noise in the NIS junction ($\bar{\Delta}_T^{NIS}$) as a function of $\Delta T$, and compare it to that of the NIN junction ($\bar{\Delta}_T^{NIN}$), as shown in the insets of Fig. \ref{fig102}. We find that both $\bar{\Delta}_T^{NIS}$ and $\bar{\Delta}_T^{NIN}$ increase quadratically with $\Delta T$ in the transparent ($Z \to 0$), intermediate ($Z = 1$), and tunneling ($Z = 5$) regimes. While the qualitative behavior of both quantities with respect to temperature bias is similar, the magnitude of $\bar{\Delta}_T^{NIS}$ is significantly larger than that of $\bar{\Delta}_T^{NIN}$, see Fig. \ref{fig103}(b) and Table \ref{Table2}. Specifically, in the transparent limit, $\bar{\Delta}_T^{NIS}$ is approximately 8 times larger than $\bar{\Delta}_T^{NIN}$; in the intermediate regime ($Z = 1$), it is roughly 4 times larger; and in the tunneling regime ($Z = 5$), about 2 times larger—see Fig. \ref{fig103}(b) and Table \ref{Table2}.} This observation further underscores the distinct noise characteristics of the NIS junction as a function of barrier strength and temperature bias.

{Our work is the first to systematically study the dependence of $\Delta_T$ noise on temperature bias ($\Delta T$) in NIS junction, a key aspect that has never been addressed before. This contribution alone is a major advancement in the understanding of noise in hybrid mesoscopic junctions.}

%{In contrast, in our work, we explicitly calculate and analyze the quantum shot noise at zero bias voltage but finite temperature bias, which is the \(\Delta_T\) noise. Unlike the total quantum noise, the quantum shot noise contribution at zero voltage bias but at finite temperature bias, i.e., $\Delta_T$ noise offers a more refined understanding of the fundamental processes at play, particularly in the presence of thermal gradients and varying barrier strengths. Furthermore, by comparing the quantum shot noise behavior in NIS and NIN junctions, we uncover unique trends and dependencies that have not been previously explored, thereby providing a more comprehensive characterization of \(\Delta_T\) noise in mesoscopic systems. This additional layer of analysis sets our work apart and fills a crucial gap left unaddressed in the existing literature.}

\section{Experimental realization and Conclusion}
\label{conclusion}

{In recent studies, $\Delta_T$ noise has been measured at a finite temperature difference in atomic scale molecular setups \cite{atomicscaleexpt,qcircuitexpt,tjunctionexpt}. In Ref.~\cite{atomicscaleexpt}, $\Delta_T$ noise is measured in an atomic-scale molecular junction due to a finite temperature gradient at zero bias voltage. In our setup, we have a metal/insulator/$s$-wave superconductor junction. For BCS superconductors, the ratio $\Delta_0/ E_F$ is significantly small \cite{Tc, Tc1, heatcurrentexpt, NIS}, for example, in the case of Nb$_3$Sn, with $\Delta_0 = 1.76 k_B T_c$ at $T_c = 18K$ \cite{Fano1, Fano}, $\Delta_0 / E_F \approx 10^{-2}-10^{-3}$, which indicates the validity of Andreev approximation.}

{{
The transition between the transparent (metallic) to the tunneling regime can be achieved experimentally in a normal metal-insulator-superconductor junction \cite{blonder}. This transition was first analyzed theoretically in Ref.~\cite{BTK}, where it is shown that the current--voltage ($I$--$V$) characteristics are curved in a normal metal--insulator--superconductor (N--I--S) junction across all the interfacial barrier strengths ($Z$) as a result of Andreev reflection in contrast to the linear $I$--$V$ relation in a normal metal--insulator--normal metal (N--I--N) junction. Ref.~\cite{blonder} experimentally verified these results using Copper (Cu) as the normal metal and Niobium (Nb) as the superconductor. In this context, $Z$ serves as a phenomenological parameter for interfacial scattering due to oxides, defects, or surface roughness. The Oxide patches are formed naturally whenever a clean metal or superconductor surface is exposed to air. $Z$ is related to the measured resistance by $R = \tfrac{R_0}{2}(1 + Z^2)$, with $R_0 = h/2e^2$. The extra factor of 2 ensures that when $Z \to 0$, $R$ approaches $R_0/2$ due to perfect Andreev reflection.

Experimentally, the NIS junction is realized by pressing the sharpened tip of a superconducting wire against a metallic disk. Prior to making contact, the metallic disk is mechanically polished first to achieve a mirror-like smoothness and then chemically polished to produce a clean, but slightly rough surface, which stabilizes the contact and allows more variation in $Z$. The superconducting tip is prepared similarly, and during the preparation process both materials inevitably develop thin oxide layers when exposed to air. As discussed in Ref.~\cite{blonder}, the copper oxide that is formed on the metallic disc made up of Copper is semiconducting, while the niobium oxide that is formed on the surface of Niobium is more complicated, containing a metallic but disordered layer. These oxide layers, which are typically tens of angstroms thick, are unavoidably present before the junction is formed. These oxides initially dominate the interface, so that when the superconducting tip first makes contact with the metallic surface, the measured resistance is usually very high, and the $I$--$V$ characteristics are linear rather than curved, acting as a N-I-N junction. This happens because the junction effectively becomes Normal metal--semiconducting oxide--dirty metallic oxide--superconductor junction, where the oxides suppress the proximity effect of the superconductor producing a linear $I$--$V$ characteristics. This corresponds to the tunneling regime with large $Z$. However, the oxide layers can be gradually scrapped by the mechanical action of pressing and sliding the superconducting tip against the normal metallic disk, increasing the transparency of the interface. As this happens, the measured $I$--$V$ characteristics evolve from being linear to curved. Ref.~\cite{blonder} demonstrates that with repeated tip movement, the experimental $I$--$V$ curves not only reproduce the qualitative trends predicted in Ref.~\cite{BTK}, but also achieve excellent quantitative agreement. The resistance decreases progressively as the oxides are removed, ultimately reaching a minimum value corresponding to the transparent limit. Once the resistance is measured in this state, the barrier strength $Z$ can be reliably extracted. Thus, by mechanically controlling the contact between the superconducting tip and the metallic disk, one can tune the junction continuously from the tunneling regime to the transparent regime.  }}

{In our setup, $\Delta_T$ noise is measured at vanishing net charge current. The $\Delta_T$ noise contribution can be isolated in principle by subtracting the total quantum noise measured at $\Delta T = 0$, which is basically the thermal noise-like contribution, from the total quantum noise measured at finite $\Delta T$ \cite{atomicscaleexpt, qcircuitexpt, tjunctionexpt, Prokudina2024}. The subtraction provides a direct measure of the predicted $\Delta_T$ noise.} To measure $\Delta_T$ noise, the normal metal is connected to a heated diffusive wire, such that the normal metal is at a higher temperature than the superconductor. A finite current passes through the wire to control its temperature, thereby maintaining a temperature difference between both junction contacts, as mentioned in Ref.~\cite{tjunctionexpt}. The current is applied only to the heated wire, while no current is applied in the junction, ensuring that no charge current but a pure heat current flows through the junction, isolating $\Delta_T$ noise solely due to the temperature difference \cite{tjunctionexpt}. This experimental technique can measure $\Delta_T$ noise in a normal metal/insulator/superconductor junction.

% {To design a NIS junction, a thin copper (Cu) film can be deposited (sputtered or patterned via lithography) on a Niobium-Tin (Nb$_3$Sn) superconductor, which can be a suitable normal metal/superconductor sample \cite{Fano}. To create an insulating barrier between the metal and the superconductor, the metal surface can be oxidized by ion beam lithography \cite{NIS}. To measure $\Delta_T$ noise, the normal metal is connected to a heated diffusive wire, such that the normal metal is at a higher temperature than the superconductor. A finite current passes through the wire to control its temperature, thereby maintaining a temperature difference between both junction contacts, as mentioned in Ref.~\cite{tjunctionexpt}. The current is applied only to the heated wire, while no current is applied in the junction, ensuring that no charge current but a pure heat current flows through the junction, isolating $\Delta_T$ noise solely due to the temperature difference \cite{tjunctionexpt}. This experimental technique can measure $\Delta_T$ noise in a normal metal/insulator/superconductor junction.}

{Our investigation delves into the behavior of \(\Delta_T\) noise in NIS junction, and we compare it with that of NIN junction to uncover key distinctions. Our analysis of \(\Delta_T\) noise reveals distinct trends: at a fixed temperature bias, \(\Delta_T\) noise in a NIS junction is zero in the transparent limit, increases with \(Z\), peaks at an intermediate value of \(Z\), and then declines as \(Z\) increases further. Notably, \(\Delta_T\) noise in the NIS junction is substantially higher (around 16 times) than that of the NIN junction in the transparent limit, but it decreases more rapidly with increasing \(Z\). {Similarly, the normalized $\Delta_T$ noise akin to the ``Fano factor" in NIS junction is also significantly higher (approximately 8 times) than that of NIN junction in the transparent regime and decreases with increasing $Z$, but even in tunneling regime, it is around 2, while the unnormalized $\Delta_T$ noise ratio almost vanish in tunneling regime.} These findings underscore the profound impact of Andreev reflection on the noise characteristics of these systems, offering valuable insights into the interplay between barrier strength, temperature bias, and noise behavior in NIS and NIN junctions.}

\section*{Appendix}

The Appendix is divided into two sections. First, in Appendix \ref{App_I}, we calculate the charge current in normal metal contact in NISand NIN junctions. Subsequently, we derive the general expression for quantum noise auto-correlation in Appendix \ref{App_ch_Qn} for NIN and NIS junctions. 
% Finally, we add the Mathematica code to calculate charge $\Delta_T$ noise in SM \cite{suppl}.

\appendix
\label{appendix}

\begin{widetext}

\section{Average charge current}
\label{App_I}

The average charge current ($\langle I_i \rangle$) in the normal metal contact ($i=1$) in a NIS junction ~\cite{BTK,lambert},

\begin{eqnarray}
\langle I_1 \rangle= \frac{2e}{h} \sum_{\substack{k,l \in \{1,2\}, \\ \rho,\gamma,\eta \in \{e,h\}}} sgn(\rho) \int^{\infty}_{0}dE A_{k\gamma;l \eta}(1 \rho,E) \langle a^{\dagger}_{k \gamma}(E) a_{l \eta}(E) \rangle,
\label{App_I1}
\end{eqnarray}
where $A_{k\gamma;l \eta}(1 \rho,E) = \delta_{1k} \delta_{1l} \delta_{\rho \gamma} \delta_{\rho \eta} - s^{\rho \gamma \dagger}_{1k} s^{\rho \eta}_{1l}$. The expectation value of $ a^{\dagger}_{k \gamma}(E) a_{l \eta}(E) \neq 0 $ only when $k=l$ and $\gamma=\eta$ and it is denoted as Fermi function in contact $k$ for particle $\gamma$ (see Ref.~\cite{datta}), i.e., $\langle a^{\dagger}_{k \gamma}(E) a_{l \eta}(E) \rangle = \delta_{kl} \delta_{\gamma \eta} f_{k \gamma}(E)=f_{l \eta}(E)$. Fermi function in contact $k \in \{1,2\}$, for particle $\gamma \in \{e,h \}$ is $f_{k \gamma}(E) = \left[ 1+ e^{\frac{E+ sgn(\gamma) V_{k}}{k_B T_{k}}} \right]^{-1}$, and $sgn(\rho)= sgn(\gamma)=sgn(\eta)=+1$ for electron and $-1$ for hole. Thus, charge current in normal metal in a NIS junction given in Eq.~(\ref{App_I1}) can be further simplified as:

\begin{eqnarray}
\langle I^{NIS}_1\rangle &=& \frac{2e}{h} \int^{\infty}_{0}\sum_{\substack{k,l \in \{1,2\}, \\ \rho, \gamma, \eta \in \{e,h\}}} sgn(\rho) \delta_{k l} \delta_{\gamma \eta} f_{k \gamma} A_{k\gamma;l \eta}(1 \rho,E) dE = \frac{2e}{h} \int^{\infty}_{0}\sum_{\substack{l \in \{1,2\}, \\ \rho, \eta \in \{e,h\}}} sgn(\rho) \left( \delta_{1l} \delta_{\rho \eta} - | s^{\rho \eta}_{1 l} |^2 \right) f_{l \eta} dE .
\end{eqnarray}
Fermi function of electron-like and hole-like quasiparticles are same in the superconductor (with $V_2=0$) and can be expressed as $f_{2e} = f_{2h} = \left( 1 + e^{\frac{E}{k_B T_2}} \right)^{-1}$ \cite{datta,noise}. Similarly, for the normal metal, the Fermi functions for electron and hole (with $V_1=eV$) are given as $f_{1e} = \left( 1 + e^{\frac{E - eV}{k_B T_1}} \right)^{-1}$ and $f_{1h} = \left( 1 + e^{\frac{E + eV}{k_B T_1}} \right)^{-1}$. {Subtracting $f_{1h}(E)$ from 1, we find $
1 - f_{1h}(E) = 1 - \frac{1}{1 + e^{\frac{E + eV}{k_B T_1}}} = \frac{e^{\frac{E + eV}{k_B T_1}}}{1 + e^{\frac{E + eV}{k_B T_1}}}.
$
Simplifying further, this becomes $
1 - f_{1h}(E) = \frac{1}{1 + e^{\frac{-E - eV}{k_B T_1}}} = f_{1e}(-E).
$}
Using the properties $f_{1e}(-E)= 1- f_{1h}(E)$, and $f_{2e}(-E)= 1- f_{2h}(E)$, we get the average charge current in normal metal in a NIS junction (see, Refs.~\cite{BTK,lambert}) as follows,
\begin{eqnarray}
\langle I^{NIS}_1 \rangle &=& \frac{2e}{h} \int^{\infty}_{-\infty} ( 1 + |s^{he}_{11}|^2 - |s^{ee}_{11}|^2 ) ( f_{1e} - f_{2e}) dE = \frac{2e}{h} \int^{\infty}_{-\infty} \mathcal{T}_{NIS} ( f_{1e} - f_{2e}) dE,
\label{A_I}
\end{eqnarray}
where $F^{NIS}_I=1+\mathcal{A}-\mathcal{B}$, with $\mathcal{A}=|s^{he}_{11}|^2$ being the Andreev reflection probability and $\mathcal{B}=|s^{ee}_{11}|^2$ is the normal reflection probability. To obtain the simplified current expression as given in Eq. (\ref{A_I}), we have used the properties $\mathcal{A}(-E)=\mathcal{A}$, and $\mathcal{B}(-E)=\mathcal{B}$ \cite{BTK,lambert, noise}, along with the properties of probability conservation, i.e., $\mathcal{A}+\mathcal{B}+\mathcal{C}+\mathcal{D}=1$, where $\mathcal{C}$ and $\mathcal{D}$ denote electron-like and hole-like transmission probabilities. In our work, we consider $T_1 = \bar{T} + \Delta T/2$ and $T_2 = \bar{T} - \Delta T/2$, where $\bar{T}$ is the average temperature of the setup, whereas $\Delta T$ is the temperature bias. The conductance in NIS junction is $G_{NIS} = \frac{4e^2}{h} \int_{-\infty}^{\infty}dE \mathcal{T}_{NIS} \left(-\frac{df}{dE}\right)$, where $f = \left(1 + e^{\frac{E}{k_B \bar{T}}}\right)^{-1}$ \cite{BTK}.

For a NIN junction, average charge current is $I^{NIN}_1 = \frac{2e}{h} \int^{\infty}_{-\infty} dE \,\,\, \mathcal{T}_{NIN} (f_{1e} -f_{2e} )$, where $\mathcal{T}_{NIN}$ is the net transmission probability through the NIN junction. Similarly, the conductance in NIN junction is given as $G_{NIN} = \frac{4e^2}{h} \int_{-\infty}^{\infty}dE \mathcal{T}_{NIN} \left(-\frac{df}{dE}\right)$, where $f = \left(1 + e^{\frac{E}{k_B \bar{T}}}\right)^{-1}$ \cite{BTK}.

\section{Quantum noise}
\label{App_ch_Qn}

The charge current-current correlation~\cite{noise, thermalnoise} between contacts $p$ and $q$ is defined $Q_{p q} (t-t^{\prime}) = \frac{1}{2 \pi} \langle \Delta I_{p}(t) \Delta I_{q}(t^{\prime}) + \Delta I_{q}(t^{\prime}) \Delta I_{p}(t) \rangle$, with $\Delta I_{p} = I_{p} - \langle I_{p} \rangle$, where $I_{p}$ is current in lead $p$. On Fourier transforming, we obtain the quantum noise power, expressed in terms of frequency as $ 2 \pi \delta(\omega + \bar{\omega}) Q_{p q}(\omega) \equiv \langle \Delta I_{p}(\omega) \Delta I_{q}(\bar{\omega}) + \Delta I_{q}(\bar{\omega}) \Delta I_{p}(\omega) \rangle $. Quantum noise at zero frequency $Q_{p q}(\omega=\bar{\omega} =0$) \cite{datta}, in a NIS junction is given as,
\begin{equation}
Q^{NIS}_{pq}  = \sum_{x, y \in \{e, h \}} Q_{pq}^{xy} =   \frac{2e^2}{h} \int \sum_{ \substack{k,l \in \{1, 2\} ,\\
x,y,\gamma,\eta \in \{e,h\}} } sgn(y) sgn(x) A_{l,\eta;k,\gamma}(q y,E) A_{k,\gamma;l,\eta}(p x,E)( \textit{f}_{k \gamma}(E) [1-\textit{f}_{l \eta}(E)] + \textit{f}_{l \eta}(E) [1-\textit{f}_{k \gamma}(E)]) dE,
\label{eqn:sn}
\end{equation}{}
where $A_{k,\gamma;l,\eta}(p x,E) = \delta_{p k} \delta_{p l} \delta_{x \gamma} \delta_{x \eta} - s^{x \gamma *}_{p k}(E) s^{x \eta}_{p l}(E)$. Here $sgn(x)=sgn(y)=+1$ for electron and $sgn(x)=sgn(y)=-1$ for hole.\\

%----------------------------------------------------------

{In a NIS junction, the scattering matrix ($S$) describes how the incoming electron and hole from the normal metal side are scattered into the outgoing electron and hole in both the normal metal and the superconductor. This $S$ matrix captures normal reflection, Andreev reflection, and transmission processes, which are essential to understanding the electronic transport properties in a NIS junction.}

{The scattering matrix $ S_{NIS} $ relating the incoming and outgoing states in a NIS junction, is given as,
\begin{eqnarray}
\begin{pmatrix}
c^-_{1e} \\
c^+_{2h} \\
c^+_{1h} \\
c^-_{2h}
\end{pmatrix}
=
S_{NIS}
\begin{pmatrix}
c^+_{1e} \\
c^-_{2e} \\
c^-_{1h} \\
c^+_{2h}
\end{pmatrix},
\text{with} \,\,\,\,\, S_{NIS} = 
\begin{pmatrix}
s^{ee}_{11} & s^{ee}_{12} & s^{eh}_{11} & s^{eh}_{12} \\
s^{ee}_{21} & s^{ee}_{22} & s^{eh}_{21} & s^{eh}_{22} \\
s^{he}_{11} & s^{he}_{12} & s^{hh}_{11} & s^{hh}_{12} \\
s^{he}_{21} & s^{he}_{22} & s^{hh}_{21} & s^{hh}_{22}
\end{pmatrix} 
\label{c1:S_mat_NIS}
\end{eqnarray}}

{The incoming and outgoing electron/hole states are denoted as follows, incoming electron in the normal metal as $ c^+_{1e} $, outgoing electron in the normal metal as $ c^-_{1e} $, incoming hole in the normal metal as $ c^-_{1h} $, outgoing hole in the normal metal as $ c^+_{1h} $, incoming electron in the superconductor as $ c^-_{2e} $, outgoing electron in the superconductor as $ c^+_{2e} $, incoming hole in the superconductor, $ c^+_{2h} $, outgoing hole in the superconductor, $ c^-_{2h} $.}

% {The incoming and outgoing states are related by a $4 \times 4$ scattering matrix such as $c_{out} = S_{NIS} c_{in}$, where the outgoing state is $c_{out} = (c^-_{1e},s^+_{2e},c^+_{1h},s^-_{2h})^T$ and the incoming state is $c_{in} = (c^+_{1e},s^-_{2e},c^-_{1h},s^+_{2h})^T$.}

{Here, the elements of the scattering matrix represent the normal and Andreev reflection and transmission amplitudes for a NIS junction. For electron incident from normal metal, $s^{ee}_{11}=b$ represents normal reflection of an incident electron reflecting as an electron, $s^{he}_{11}=a$ is Andreev reflection amplitude of an electron reflecting as a hole, $s^{ee}_{21}=\sqrt[]{|u|^2- |v|^2} c$ represents transmission amplitude of an incident electron scattered as an electron, $s^{he}_{21}=\sqrt[]{|u|^2-|v|^2}d$ is transmission amplitude of an incident electron scattered as a hole. For a NIS junction, the scattering amplitudes $a, b, c$ and $d$ have been derived in Ref. \cite{BTK}, which are given as,}

{
\begin{align}
a &= \frac{u v}{\alpha}, \quad
b = -\frac{(u - v)(Z^2 + iZ)}{\alpha}, \quad
c = \frac{u(1 - iZ)}{\alpha}, \quad
d = \frac{i v Z}{\alpha}, \,\,\,\, \text{where} \,\,\,\,\,\, \alpha = u^2 + (u^2 - v^2) Z^2,
\end{align}}

{where
$Z$ is the barrier strength. The coherence factors for energy $E$ above the superconducting gap $\Delta_0$ are $u(v) = \left[ \frac{1}{2} \left\{ 1 \pm \frac{ \sqrt[]{ E^2 - \Delta^2_0 }}{E} \right\} \right]^{1/2}$ and below $\Delta_0$ are $u(v) = \left[ \frac{1}{2} \left\{ 1 \pm \frac{i \sqrt[]{ \Delta_0^2 - E^2 }}{E} \right\} \right]^{1/2}$.}

{Similarly, $s^{ee}_{22}$, $s^{eh}_{22} $, $s^{he}_{22}$, $s^{hh}_{22}$ are normal and Andreev reflections within the superconductor, while $s^{ee}_{12}$, $s^{eh}_{12}$, $s^{he}_{12}$, $s^{hh}_{12}$ denote  transmissions amplitudes from superconductor to normal metal. The Andreev reflection probability $\mathcal{A}$ is denoted as $|s_{11}^{he}|^2 = |a|^2$, normal reflection probability $\mathcal{B}$ is denoted as $|s_{11}^{ee}|^2 = |b|^2$, transmission probability from normal metal to superconductor as an electron-like quasiparticle is denoted as $\mathcal{C} = |s_{21}^{ee}|^2 = (|u|^2 - |v|^2)|c|^2$, transmission probabibility from normal metal to superconductor as a hole like quasiparticle is $\mathcal{D} = |s_{21}^{he}|^2 = (|u|^2 - |v|^2)|d|^2$.}

{From unitarity of $S_{NIS}$ see Eq. (3), we get :  $s^{ee}_{12} =  s^{ee}_{21}; ~  s^{ee}_{22}= s^{ee}_{11}; s^{eh}_{11}= s^{he}_{11}; ~ s^{eh}_{21}= s^{he *}_{21}; 
     s^{he}_{12}= -s^{he}_{21}; ~ s^{he}_{22}= -s^{he}_{11}; s^{hh}_{11}= s^{ee *}_{11};~  s^{hh}_{21}= s^{ee *}_{21}; 
     s^{eh}_{12}= -s^{he *}_{21};  s^{hh}_{12}= s^{ee *}_{21}; s^{eh}_{22}= -s^{he}_{11}; \text{and} ~  s^{hh}_{22}= s^{ee *}_{11}$. }
     
% \begin{eqnarray}
%     s^{ee}_{12} =  s^{ee}_{21};~  s^{ee}_{22}= s^{ee}_{11}; &&~  s^{eh}_{11}= s^{he}_{11}; ~ s^{eh}_{21}= s^{he *}_{21}; \nonumber \\
%      s^{he}_{12}= -s^{he}_{21}; ~ s^{he}_{22}= -s^{he}_{11}; && ~ s^{hh}_{11}= s^{ee *}_{11};~  s^{hh}_{21}= s^{ee *}_{21}; \nonumber \\
%      s^{eh}_{12}= -s^{he *}_{21}; ~ s^{hh}_{12}= s^{ee *}_{21}; && ~ s^{eh}_{22}= -s^{he}_{11};~  s^{hh}_{22}= s^{ee *}_{11}.
%      \label{S_rel}
% \end{eqnarray}
% }

{Thus, the $4 \times 4$, $s$-matrix $ S_{NIS} $ can be written as,
\begin{align}
S_{NIS} =
\begin{pmatrix}
b & \sqrt{|u|^2 - |v|^2}c & a & -\sqrt{|u|^2 - |v|^2} d^* \\
\sqrt{|u|^2 - |v|^2}c & b & \sqrt{|u|^2 - |v|^2}d^* & -a \\
a & -\sqrt{|u|^2 - |v|^2}d & b^* & \sqrt{|u|^2 - |v|^2}c^* \\
\sqrt{|u|^2 - |v|^2}d & -a & \sqrt{u^2 - v^2}c^* & b^*
\end{pmatrix}
\label{c1:S_mat1_NIS}
\end{align}
}

{Furthermore, scattering amplitudes in Eq. (5) obey the following relations, from unitarity of $S_{NIS}$,
\begin{equation}
a b = (|u|^2 - |v|^2) c d^*, \quad a b^* = (|u|^2 - |v|^2) c^* d.
\label{S_rel1}
\end{equation}}

{This relation leads to another relation, i.e., $\mathcal{A} \mathcal{B} = \mathcal{C} \mathcal{D}$.}
{Quantum noise auto-correlation (i.e., $p=q=1$) in the normal metal of a NIS junction is
\begin{eqnarray}
Q^{NIS}_{11} && = \frac{2e^2}{h} \int \sum_{ \substack{k,l \in \{1, 2\} ,\\ x,y,\gamma,\eta \in \{e,h\}} } sgn(y) sgn(x) A_{l,\eta;k,\gamma}(1y,E) A_{k,\gamma;l,\eta}(1 x,E)( \textit{f}_{k \gamma}(E) [1-\textit{f}_{l \eta}(E)] + \textit{f}_{l \eta}(E) [1-\textit{f}_{k \gamma}(E)]) dE.
\label{eq:QNIS}
\end{eqnarray}}

{
Fermi function of electron-like and hole-like quasiparticles are same in the superconductor (with $V_2=0$) and can be expressed as $f_{2e} = f_{2h} = \left( 1 + e^{\frac{E}{k_B T_2}} \right)^{-1}$ \cite{datta,noise}. Similarly, for the normal metal, the Fermi functions for electron and hole (with $V_1=eV$) are given as $f_{1e} = \left( 1 + e^{\frac{E - eV}{k_B T_1}} \right)^{-1}$ and $f_{1h} = \left( 1 + e^{\frac{E + eV}{k_B T_1}} \right)^{-1}$.}

{Considering both electron and hole, quantum noise auto-correlation ($Q^{NIS}_{11}$) can be separated as current-current correlations of the same particles ($Q^{sa}_{11}$) and current-current correlations of different particles ($Q^{op}_{11}$), i.e., $Q^{NIS}_{11} = Q^{sa}_{11} + Q^{op}_{11} = (Q^{ee}_{11} + Q^{hh}_{11}) + (Q^{eh}_{11} + Q^{he}_{11})$, see Ref.~\cite{datta}, where $Q^{sa}_{11} = Q^{ee}_{11} + Q^{hh}_{11}= \langle \Delta \hat{I}_{1e} \Delta \hat{I}_{1e} \rangle + \langle \Delta \hat{I}_{1h} \Delta \hat{I}_{1h} \rangle$, and $Q^{op}_{11} =Q^{eh}_{11} + Q^{he}_{11} = \langle \Delta \hat{I}_{1e} \Delta \hat{I}_{1h} \rangle + \langle \Delta \hat{I}_{1h} \Delta \hat{I}_{1e} \rangle$. First, we calculate quantum noise auto-correlation for same particles, for electron ($\langle \Delta \hat{I}_{1e} \Delta \hat{I}_{1e} \rangle$) can be derived from, 
\begin{eqnarray}
Q^{ee}_{11} &=& \langle \Delta \hat{I}_{1e} \Delta \hat{I}_{1e} \rangle = \frac{2e^2}{h}  \sum_{ \substack{\{k,l\} \in \{1, 2\} ,\\ \{\gamma,\eta\} \in \{e,h\}} } \int_0^{\infty}  A_{k,\gamma;l,\eta}(1e) A_{l,\eta;k,\gamma}(1e) (\textit{f}_{k \gamma} [1-\textit{f}_{l \eta}] + \textit{f}_{l \eta} [1-\textit{f}_{k \gamma}]) dE ,
\label{c1:Qsa_ee}
\end{eqnarray}
where  $A_{k,\gamma;l,\eta}(1e)= \delta_{1k} \delta_{1l} \delta_{e\gamma} \delta_{e \eta} - s^{e \gamma \dagger}_{1k} s^{e \eta}_{1l} $ and  $ A_{l,\eta;k,\gamma}(1e) = \delta_{1l} \delta_{1k} \delta_{e\eta} \delta_{e \gamma} - s^{e \eta \dagger}_{1l} s^{e \gamma}_{1k} $. Using the scattering matrix relations given in Eq. (\ref{c1:S_mat1_NIS}), $Q^{ee}_{11}$ can be further simplified as
\begin{eqnarray}
Q^{ee}_{11} &=& \frac{2e^2}{h} \sum_{ \substack{\{k,l\} \in \{1, 2\} ,\\ \{\gamma,\eta\} \in \{e,h\}} } \int_0^{\infty} ( \delta_{1k} \delta_{1l} \delta_{e\gamma} \delta_{e \eta} - s^{e \gamma \dagger}_{1k} s^{e \eta}_{1l} ) (\delta_{1l} \delta_{1k} \delta_{e\eta} \delta_{e \gamma} - s^{e \eta \dagger}_{1l} s^{e \gamma}_{1k} ) (\textit{f}_{k \gamma} [1-\textit{f}_{l \eta}] + \textit{f}_{l \eta} [1-\textit{f}_{k \gamma}]) dE  \nonumber \\
&=& \frac{4e^2}{h} \int_0^{\infty} \biggl[ (1 - \mathcal{B})^2 f_{1e} (1-f_{1e}) + 
\mathcal{A} \mathcal{B}~ f_{1e} (1-f_{1h}) + \mathcal{A} \mathcal{B}~ f_{1h} (1-f_{1e}) +
\mathcal{A}^2 f_{1h} (1-f_{1h}) + \mathcal{B}\mathcal{C}~ f_{1e} (1-f_{2e}) \nonumber \\
&&  + \mathcal{B}\mathcal{D}~ f_{1e} (1-f_{2e}) + \mathcal{A} \mathcal{C}~ f_{1h} (1-f_{2e}) +\mathcal{A} \mathcal{D}~ f_{1h} (1-f_{2e}) + \mathcal{B}\mathcal{C}~ f_{2e} (1-f_{1e}) + \mathcal{B}\mathcal{D}~ f_{2e} (1-f_{1e}) \nonumber \\
&&  + \mathcal{A} \mathcal{C}~ f_{2e} (1-f_{1h}) +\mathcal{A} \mathcal{D}~ f_{2e} (1-f_{1h}) + \mathcal{C}^2~ f_{2e} (1-f_{2e}) + \mathcal{D}^2~ f_{2e} (1-f_{2e}) + 2 \mathcal{C} \mathcal{D}~ f_{2e} (1-f_{2e}) \biggr] dE  \nonumber \\
&=& \frac{4e^2}{h} \int_0^{\infty} \biggl[ (1 - \mathcal{B})^2 f_{1e} (1-f_{1e}) + 
\mathcal{A} \mathcal{B}~ f_{1e} (1-f_{1h}) + \mathcal{A} \mathcal{B}~ f_{1h} (1-f_{1e}) +
\mathcal{A}^2 f_{1h} (1-f_{1h}) + \mathcal{B}(\mathcal{C}+\mathcal{D}) f_{1e} (1-f_{2e}) \nonumber \\
&&   + \mathcal{A} ( \mathcal{C}+ \mathcal{D})~ f_{1h} (1-f_{2e}) + \mathcal{B} (\mathcal{C}+ \mathcal{D})~ f_{2e} (1-f_{1e}) + \mathcal{A} (\mathcal{C} + \mathcal{D} ) ~ f_{2e} (1-f_{1h}) + ( \mathcal{C}+ \mathcal{D})^2~ f_{2e} (1-f_{2e}) \biggr] dE.
\label{c1:Qsa_ee1}
\end{eqnarray}}
 
{Similarly, using the scattering matrix relations given in Eq. (\ref{c1:S_mat1_NIS}), quantum noise auto-correlation for different particles ($Q^{eh}_{11} = \langle \Delta \hat{I}_{1e} \Delta \hat{I}_{1h} \rangle$) can be simplified as
\begin{eqnarray}
Q^{eh}_{11} &=& \langle \Delta \hat{I}_{1e} \Delta \hat{I}_{1h} \rangle = -\frac{2e^2}{h}  \sum_{\substack{\{k,l\} \in \{1, 2\} \\ \{\gamma,\eta\} \in \{e,h\}}} \int_0^{\infty}
A_{k,\gamma;l,\eta}(1e) A_{l,\eta;k,\gamma}(1h) (\textit{f}_{k \gamma} [1-\textit{f}_{l \eta}] + \textit{f}_{l \eta} [1-\textit{f}_{k \gamma}]) dE \nonumber \\
&=& -\frac{2e^2}{h} \sum_{\substack{\{k,l\} \in \{1, 2\} \\ \{\gamma,\eta\} \in \{e,h\}}} 
\int_0^{\infty} \bigl( \delta_{1k} \delta_{1l} \delta_{e\gamma} \delta_{e \eta} - s^{e \gamma \dagger}_{1k} s^{e \eta}_{1l} \bigr) \bigl( \delta_{1l} \delta_{1k} \delta_{h \eta} \delta_{h \gamma} - s^{h \eta \dagger}_{1l} s^{h \gamma}_{1k} \bigr) \times (\textit{f}_{k \gamma} [1-\textit{f}_{l \eta}] + \textit{f}_{l \eta} [1-\textit{f}_{k \gamma}]) dE, \nonumber \\
&=& \frac{4e^2}{h} \int_{0}^{\infty} \biggl[ \mathcal{A} (1 - \mathcal{B}) f_{1e} (1-f_{1e}) - 
\mathcal{A} \mathcal{B}~ f_{1e} (1-f_{1h}) - \mathcal{A} \mathcal{B}~ f_{1h} (1-f_{1e}) +
\mathcal{A}(1-\mathcal{B}) f_{1h} (1-f_{1h}) \nonumber \\
&& + (|u|^2 - |v|^2) a b^* c d^*~ f_{1e} (1-f_{2e})  + (|u|^2 - |v|^2) a b^* c d^*~ f_{1e} (1-f_{2e}) + (|u|^2 - |v|^2) a b^* c d^*~ f_{1h} (1-f_{2e}) \nonumber \\
&& + (|u|^2 - |v|^2) a b^* c d^*~ f_{1h} (1-f_{2e})  + (|u|^2 - |v|^2) a b c^* d~ f_{2e} (1-f_{1e}) + (|u|^2 - |v|^2) a b c^* d~ f_{2e} (1-f_{1e}) 
\nonumber \\ &&
+ (|u|^2 - |v|^2) a b c^* d~ f_{2e} (1-f_{1h})   + (|u|^2 - |v|^2) a b c^* d~ f_{2e} (1-f_{1h}) - 4 \mathcal{C} \mathcal{D}~ f_{2e} (1-f_{2e}) \biggr] dE . 
\label{c1:Qop_eh}
\end{eqnarray}}
{Now using the properties as shown above Eq. (\ref{eq:QNIS}), we get $(|u|^2 - |v|^2) a b^* c d^* = |a|^2 |b|^2 = \mathcal{A} \mathcal{B}, (|u|^2 - |v|^2) a b c^* d^* = |a|^2 |b|^2 = \mathcal{A} \mathcal{B}$.} 
     
{Therefore, $Q^{eh}_{11}$ can be further simplified as
\begin{eqnarray}
Q^{eh}_{11} &=& \frac{2e^2}{h} \int_0^{\infty} \biggl[ \mathcal{A} (1 - \mathcal{B}) f_{1e} (1-f_{1e}) - 
\mathcal{A} \mathcal{B}~ f_{1e} (1-f_{1h}) - \mathcal{A} \mathcal{B}~ f_{1h} (1-f_{1e}) +
\mathcal{A}(1-\mathcal{B}) f_{1h} (1-f_{1h})  \nonumber \\
&& + 2 \mathcal{A} \mathcal{B}~ f_{1e} (1-f_{2e})  + 2 \mathcal{A} \mathcal{B}~ f_{1h} (1-f_{2e}) + 2 \mathcal{A} \mathcal{B}~ f_{2e} (1-f_{1e})  + 2 \mathcal{A} \mathcal{B}~ f_{2e} (1-f_{1h}) - 4 \mathcal{C} \mathcal{D}~ f_{2e} (1-f_{2e}) \biggr] dE . 
\label{c1:Qop_eh1}
\end{eqnarray}}

{Now, quantum noise auto-correlation for different particles ($Q^{he}_{11} = \langle \Delta \hat{I}_{1h} \Delta \hat{I}_{1e} \rangle$) can be simplified as
\begin{eqnarray}
Q^{he}_{11} &=& -\frac{2e^2}{h} \sum_{\substack{\{k,l\} \in \{1, 2\} \\ \{\gamma,\eta\} \in \{e,h\}}} \int_0^{\infty}
A_{k,\gamma;l,\eta}(1h) A_{l,\eta;k,\gamma}(1e) (\textit{f}_{k \gamma} [1-\textit{f}_{l \eta}] + \textit{f}_{l \eta} [1-\textit{f}_{k \gamma}]) dE \nonumber \\
&=& \frac{4e^2}{h} \int_0^{\infty} \biggl[ \mathcal{A} (1 - \mathcal{B}) f_{1e} (1-f_{1e}) - 
\mathcal{A} \mathcal{B}~ f_{1e} (1-f_{1h}) - \mathcal{A} \mathcal{B}~ f_{1h} (1-f_{1e}) +
\mathcal{A}(1-\mathcal{B}) f_{1h} (1-f_{1h})  \nonumber \\
&& + (|u|^2 - |v|^2) a b c^* d~ f_{1e} (1-f_{2e})  + (|u|^2 - |v|^2) a b c^* d~ f_{1e} (1-f_{2e}) + (|u|^2 - |v|^2) a b c^* d~ f_{1h} (1-f_{2e}) \nonumber \\
&& + (|u|^2 - |v|^2) a b c^* d~ f_{1h} (1-f_{2e})  + (|u|^2 - |v|^2) a b c^* d^*~ f_{2e} (1-f_{1e})  + (|u|^2 - |v|^2) a b c^* d^*~ f_{2e} (1-f_{1e}) \nonumber \\
&& + (|u|^2 - |v|^2) a b c^* d^*~ f_{2e} (1-f_{1h})   + (|u|^2 - |v|^2) a b c^* d^*~ f_{2e} (1-f_{1h}) - 4 \mathcal{C} \mathcal{D}~ f_{2e} (1-f_{2e}) \biggr] dE \nonumber \\
&=& \frac{4e^2}{h} \int \biggl[ \mathcal{A} (1 - \mathcal{B}) f_{1e} (1-f_{1e}) - 
\mathcal{A} \mathcal{B}~ f_{1e} (1-f_{1h}) - \mathcal{A} \mathcal{B}~ f_{1h} (1-f_{1e}) +
\mathcal{A}(1-\mathcal{B}) f_{1h} (1-f_{1h}) + 2 \mathcal{A} \mathcal{B}~ f_{1e} (1-f_{2e}) \nonumber \\
&&  + 2 \mathcal{A} \mathcal{B}~ f_{1h} (1-f_{2e}) + 2 \mathcal{A} \mathcal{B}~ f_{2e} (1-f_{1e})  + 2 \mathcal{A} \mathcal{B}~ f_{2e} (1-f_{1h}) - 4 \mathcal{C} \mathcal{D}~ f_{2e} (1-f_{2e}) \biggr] dE 
\label{c1:Qop_he}
\end{eqnarray}}

{Similar to quantum noise auto correlation for electron-electron, quantum noise auto-correlation for same particles, for hole ($Q^{hh}_{11}= \langle \Delta \hat{I}_{1h} \Delta \hat{I}_{1h} \rangle$) can be derived as
\begin{eqnarray}
    Q^{hh}_{11} &=& \langle \Delta \hat{I}_{1h} \Delta \hat{I}_{1h} \rangle = \frac{2e^2}{h}  \sum_{ \substack{\{k,l\} \in \{1, 2\} ,\\ \{\gamma,\eta\} \in \{e,h\}} } \int_0^{\infty}  A_{k,\gamma;l,\eta}(1h) A_{l,\eta;k,\gamma}(1h) (\textit{f}_{k \gamma} [1-\textit{f}_{l \eta}] + \textit{f}_{l \eta} [1-\textit{f}_{k \gamma}]) dE \nonumber \\
    &=& \frac{4e^2}{h} \int_0^{\infty} \biggl[ \mathcal{A}^2 f_{1e} (1-f_{1e}) + 
\mathcal{A} \mathcal{B}~ f_{1e} (1-f_{1h}) + \mathcal{A} \mathcal{B}~ f_{1h} (1-f_{1e}) +
(1-\mathcal{B}^2) f_{1h} (1-f_{1h}) + \mathcal{A}(\mathcal{C}+\mathcal{D}) f_{1e} (1-f_{2e}) \nonumber \\
&&   + \mathcal{B} ( \mathcal{C}+ \mathcal{D})~ f_{1h} (1-f_{2e}) + \mathcal{A} (\mathcal{C}+ \mathcal{D})~ f_{2e} (1-f_{1e}) + \mathcal{B} (\mathcal{C} + \mathcal{D} ) ~ f_{2e} (1-f_{1h}) + ( \mathcal{C}+ \mathcal{D})^2~ f_{2e} (1-f_{2e}) \biggr] dE.
\label{c1:Qsa_hh}
\end{eqnarray}
}

{Total quantum noise auto correlation is defined as $Q^{NIS}_{11} =Q^{ee}_{11} + Q^{eh}_{11} + Q^{he}_{11} + Q^{hh}_{11}$ and expressed as:
\begin{eqnarray}
 Q^{NIS}_{11} &=&   \frac{4e^2}{h} \int^{\infty}_0 \biggl[ \{ (1 - \mathcal{B})^2 + \mathcal{A}^2 + 2 \mathcal{A} (1-\mathcal{B}) \} f_{1e} (1-f_{1e})  + \{ (1 - \mathcal{B})^2 + \mathcal{A}^2 + 2 \mathcal{A} (1-\mathcal{B}) \} f_{1h} (1-f_{1h}) \nonumber \\
 && + \{ \mathcal{B}(\mathcal{C}+\mathcal{D}) + \mathcal{A} (\mathcal{C}+\mathcal{D}) + 4 \mathcal{A} \mathcal{B} \} f_{1e} (1-f_{2e}) + \{ \mathcal{B}(\mathcal{C}+\mathcal{D}) + \mathcal{A} (\mathcal{C}+\mathcal{D}) + 4 \mathcal{A} \mathcal{B} \}  f_{2e} (1-f_{1e}) \nonumber \\
&&  + \{ \mathcal{B}(\mathcal{C}+\mathcal{D}) + \mathcal{A} (\mathcal{C}+\mathcal{D}) + 4 \mathcal{A} \mathcal{B} \} ~ f_{1h} (1-f_{2e}) + \{ \mathcal{B}(\mathcal{C}+\mathcal{D}) + \mathcal{A} (\mathcal{C}+\mathcal{D}) + 4 \mathcal{A} \mathcal{B} \} ~ f_{2e} (1-f_{1h}) \nonumber \\ 
&& + \{ 2 ( \mathcal{C}+ \mathcal{D})^2 - 4 \mathcal{C} \mathcal{D}\}~ f_{2e} (1-f_{2e}) \biggr] dE \nonumber \\
&=& \frac{4e^2}{h} \int^{\infty}_0 \biggl[ \{ (1 - \mathcal{B} + \mathcal{A})^2 \} f_{1e} (1-f_{1e})  + \{ (1 - \mathcal{B}+\mathcal{A})^2 \} f_{1h} (1-f_{1h}) + \{ (\mathcal{A}+\mathcal{B})(\mathcal{C}+\mathcal{D}) + 4 \mathcal{A} \mathcal{B} \}  \nonumber \\
&& ( f_{1e} (1-f_{2e}) + f_{2e} (1-f_{1e}))  + \{ (\mathcal{A}+\mathcal{B})(\mathcal{C}+\mathcal{D}) + 4 \mathcal{A} \mathcal{B} \} ~ ( f_{1h} (1-f_{2e}) + f_{2e} (1-f_{1h})) +  \nonumber \\
&& \{ 2 ( \mathcal{C}+ \mathcal{D})^2 - 8 \mathcal{C} \mathcal{D}\}~ f_{2e} (1-f_{2e}) \biggr] dE.
\label{c1:Q} 
\end{eqnarray}
where $(\mathcal{A}+\mathcal{B})(\mathcal{C}+\mathcal{D}) + 4 \mathcal{A} \mathcal{B} = (\mathcal{A}+\mathcal{B})(1 - \mathcal{A} - \mathcal{B}) + 4 \mathcal{A} \mathcal{B} =\mathcal{A}(1-\mathcal{A}) + \mathcal{B}(1-\mathcal{B}) + 2 \mathcal{A} \mathcal{B} $, thus
\begin{eqnarray}
 Q^{NIS}_{11} &=& \frac{4e^2}{h} \int^{\infty}_0 \biggl[ \{ (1 - \mathcal{B} + \mathcal{A})^2 \} f_{1e} (1-f_{1e})  + \{ \mathcal{A}(1-\mathcal{A}) + \mathcal{B}(1-\mathcal{B}) + 2 \mathcal{A} \mathcal{B} \} ( f_{1e} (1-f_{2e}) + f_{2e} (1-f_{1e})) \nonumber \\
&&  + \{ ( \mathcal{C}+ \mathcal{D})^2 - 4 \mathcal{C} \mathcal{D}\}~ f_{2e} (1-f_{2e})  + \{ (1 - \mathcal{B}+\mathcal{A})^2 \} f_{1h} (1-f_{1h}) + \{ \mathcal{A}(1-\mathcal{A}) + \mathcal{B}(1-\mathcal{B}) + 2 \mathcal{A} \mathcal{B} \} \nonumber \\
&&  ~ ( f_{1h} (1-f_{2e}) + f_{2e} (1-f_{1h})) + \{  ( \mathcal{C}+ \mathcal{D})^2 - 4 \mathcal{C} \mathcal{D}\}~ f_{2e} (1-f_{2e}) \biggr] dE \nonumber \\
&=& \frac{4e^2}{h} \int^{\infty}_0 \biggl[ \{ (1 - \mathcal{B} + \mathcal{A})^2 \} f_{1e} (1-f_{1e})  + \{ \mathcal{A}(1-\mathcal{A}) + \mathcal{B}(1-\mathcal{B}) + 2 \mathcal{A} \mathcal{B} \} ( f_{1e} (1-f_{2e}) + f_{2e} (1-f_{1e})) \nonumber \\
&&  + \{ ( \mathcal{C}+ \mathcal{D})^2 - 4 \mathcal{C} \mathcal{D}\}~ f_{2e} (1-f_{2e})  \biggr] dE + \frac{4e^2}{h} \int^{\infty}_0 \biggl[ \{ (1 - \mathcal{B} + \mathcal{A})^2 \} f_{1h} (1-f_{1h})  + \{ \mathcal{A}(1-\mathcal{A}) + \mathcal{B}(1-\mathcal{B})  \nonumber \\
&& + 2 \mathcal{A} \mathcal{B} \}    ( f_{1h} (1-f_{2e}) + f_{2e} (1-f_{1h})) + \{ ( \mathcal{C}+ \mathcal{D})^2 - 4 \mathcal{C} \mathcal{D}\}~ f_{2e} (1-f_{2e})  \biggr] dE 
\label{c1:Q1} 
\end{eqnarray}}

{Now, we utilize the properties of transmission probabilities $\mathcal{\mathcal{A}}(-E)=\mathcal{\mathcal{A}}$, $\mathcal{\mathcal{B}}(-E)=\mathcal{\mathcal{B}}$, $\mathcal{\mathcal{C}}(-E)=\mathcal{\mathcal{C}}$, and $\mathcal{\mathcal{D}}(-E)=\mathcal{\mathcal{D}}$ \cite{BTK,noise,martin} and Fermi functions, i.e., $f_{1h}(-E)= 1- f_{1e}(E)$, and $f_{2e}(-E)= 1- f_{2e}(E)$ \cite{BTK,noise,lambert}, we can further rewrite $Q_{11}^{NIS}$ as}

{
\begin{eqnarray}
Q_{11}^{NIS} &=& \frac{4e^2}{h} \int^{\infty}_0 \biggl[ \{ (1 - \mathcal{B} + \mathcal{A})^2 \} f_{1e} (1-f_{1e})  + \{ \mathcal{A}(1-\mathcal{A}) + \mathcal{B}(1-\mathcal{B}) + 2 \mathcal{A} \mathcal{B} \} ( f_{1e} (1-f_{2e}) + f_{2e} (1-f_{1e}))  \nonumber \\
&&  + \{ ( \mathcal{C}+ \mathcal{D})^2 - 4 \mathcal{C} \mathcal{D}\}~ f_{2e} (1-f_{2e})  \biggr] dE \nonumber + \frac{4e^2}{h} \int^0_{-\infty} \biggl[ \{ (1 - \mathcal{B} + \mathcal{A})^2 \} f_{1e} (1-f_{1e})  + \{ \mathcal{A}(1-\mathcal{A}) + \mathcal{B}(1-\mathcal{B}) + 2 \mathcal{A} \mathcal{B} \}  \nonumber \\
&&  ( f_{1e} (1-f_{2e}) + f_{2e} (1-f_{1e})) + \{ ( \mathcal{C}+ \mathcal{D})^2 - 4 \mathcal{C} \mathcal{D}\}~ f_{2e} (1-f_{2e})  \biggr] dE \nonumber \\
&=& \frac{4e^2}{h} \int^{\infty}_{-\infty} \biggl[ \{ (1 - \mathcal{B} + \mathcal{A})^2 \} f_{1e} (1-f_{1e})  + \{ \mathcal{A}(1-\mathcal{A}) + \mathcal{B}(1-\mathcal{B}) + 2 \mathcal{A} \mathcal{B} \} ( f_{1e} (1-f_{2e}) + f_{2e} (1-f_{1e}))  \nonumber \\
&&  + \{ ( \mathcal{C}+ \mathcal{D})^2 - 4 \mathcal{C} \mathcal{D}\}~ f_{2e} (1-f_{2e})  \biggr] dE.
\label{c1:Q10} 
\end{eqnarray}}

{Thus $Q^{NIS}_{11}$ can be further simplified as
\begin{eqnarray}
 Q^{NIS}_{11} &=& \frac{4e^2}{h} \int^{\infty}_{-\infty} \biggl[ \{ (1 - \mathcal{B} + \mathcal{A})^2 \} f_{1e} (1-f_{1e}) + \{ \mathcal{A}(1-\mathcal{A}) + \mathcal{B}(1-\mathcal{B}) + 2 \mathcal{A} \mathcal{B} \} (f_{1e} - f^2_{1e} + f^2_{1e} - f_{1e} f_{2e} + f_{2e}\nonumber \\ 
 && - f^2_{2e} + f^2_{2e} - f_{1e} f_{2e} ) +  \{ ( \mathcal{C}+ \mathcal{D})^2 - 4 \mathcal{C} \mathcal{D}\}~ f_{2e} (1-f_{2e}) \biggr] dE \nonumber \\
 &=& \frac{4e^2}{h} \int^{\infty}_{-\infty} \biggl[ \{ (1 - \mathcal{B} + \mathcal{A})^2 + \mathcal{A}(1-\mathcal{A}) + \mathcal{B}(1-\mathcal{B}) + 2 \mathcal{A} \mathcal{B} \} f_{1e} (1-f_{1e}) + \{ ( \mathcal{C}+ \mathcal{D})^2 - 4 \mathcal{C} \mathcal{D} \nonumber \\ 
 && + \mathcal{A}(1-\mathcal{A}) + \mathcal{B}(1-\mathcal{B}) + 2 \mathcal{A} \mathcal{B}\}~ f_{2e} (1-f_{2e}) + \{ \mathcal{A}(1-\mathcal{A}) + \mathcal{B}(1-\mathcal{B}) + 2 \mathcal{A} \mathcal{B} \} ( f_{1e} -f_{2e})^2 \biggr] dE \nonumber \\
&=& \frac{4e^2}{h} \int^{\infty}_{-\infty} \biggl[ (1 + \mathcal{B}^2 + \mathcal{A}^2 + 2 \mathcal{A} -2 \mathcal{B} - 2 \mathcal{A}\mathcal{B} + \mathcal{A} - \mathcal{A}^2 + \mathcal{B} -\mathcal{B}^2 + 2 \mathcal{A} \mathcal{B} ) f_{1e} (1-f_{1e}) + (1 + \mathcal{B}^2 + \mathcal{A}^2 \nonumber  - 2 \mathcal{A} -2 \mathcal{B} \nonumber \\ 
 && + 2 \mathcal{A}\mathcal{B}  - 4 \mathcal{C} \mathcal{D} + \mathcal{A}  - \mathcal{A}^2 + \mathcal{B} -\mathcal{B}^2 + 2 \mathcal{A} \mathcal{B})~ f_{2e} (1-f_{2e}) + \{ \mathcal{A}(1-\mathcal{A}) + \mathcal{B}(1-\mathcal{B})   + 2 \mathcal{A} \mathcal{B} \} ( f_{1e} -f_{2e})^2 \biggr] dE \nonumber \\
&=& \frac{4e^2}{h} \int^{\infty}_{-\infty} \biggl[ (1 - \mathcal{B} + 3 \mathcal{A} ) f_{1e} (1-f_{1e}) + (1 - \mathcal{B} - \mathcal{A})~ f_{2e} (1-f_{2e}) + \{ \mathcal{A}(1-\mathcal{A}) + \mathcal{B}(1-\mathcal{B}) + 2 \mathcal{A} \mathcal{B} \} ( f_{1e} -f_{2e})^2 \biggr] dE \nonumber \\
&=& Q^{NIS}_{11th} + Q^{NIS}_{11sh}
\label{c1:Q2} 
\end{eqnarray}
}

{While getting the final expression of Eq. (\ref{c1:Q2}), we utilize the property $\mathcal{\mathcal{A} \mathcal{B}} = \mathcal{\mathcal{C} \mathcal{D}}$ in NIS junction. $Q_{11th}^{NIS}$ and $Q_{11sh}^{NIS}$ are the thermal noise-like contribution and shot noise-like contribution to the total quantum noise correlation $Q_{11}^{NIS}$. The mathematical expression for $Q_{11th}^{NIS}$ and $Q_{11sh}^{NIS}$ are given as}

{
\begin{equation}
    \begin{split}
        Q_{11th}^{NIS} &= \frac{4e^2}{h} \int^{\infty}_{-\infty} \biggl[ (1 - \mathcal{B} + 3 \mathcal{A} ) f_{1e} (1-f_{1e}) + (1 - \mathcal{B} - \mathcal{A})~ f_{2e} (1-f_{2e}) \biggl], \, \,\,\, \text{and} \\
        Q_{11sh}^{NIS} &= \frac{4e^2}{h} \int^{\infty}_{-\infty} \biggl[ \mathcal{A}(1-\mathcal{A}) + \mathcal{B}(1-\mathcal{B}) + 2 \mathcal{A} \mathcal{B}  \biggr] ( f_{1e} -f_{2e})^2 dE.
    \end{split}
    \label{B:100}
\end{equation}}

    {While it is true that single-particle transmission is suppressed for \( E < \Delta \), the {transport in the NIS junction still proceeds via Andreev reflection}, which corresponds to the effective conversion of single particle electron/hole states in normal metal to Cooper pairs inside the superconductor. It is well known that the current below the gap as shown in Eq. (\ref{A_I}) simplifies to \cite{BTK}},
    {
\begin{equation}
\langle I_1^{NIS} \rangle = \frac{2e}{h} \int dE \, 2\mathcal{A} (f_{1e} - f_{2e}).
\label{curr}
\end{equation}}
{Therefore, there is a net transport from normal metal to superconductor below the gap. This leads to a shot noise below the gap also. Accordingly, the shot noise expression from Eq. (\ref{B:100}) is,}
{
\begin{equation}
Q_{11sh}^{\mathrm{NIS}} = \frac{4e^2}{h} \int dE \, 4\mathcal{A}(1-\mathcal{A}) (f_{1e} - f_{2e})^2.
\label{nois}
\end{equation}}
{Eqs. (\ref{curr}) and (\ref{nois}) clearly demonstrate that both {shot noise and charge current are finite below the gap} and this formula is valid whether one takes voltage bias or temperature bias.}

{Now, we can derive the expression of quantum noise in NIN junction in a similar manner. In NIN junction, only the normal reflection and transmission of electron occur and there is no Andreev reflection. The $S$-matrix for the NIN junction is therefore,}

{The scattering matrix $ S_{NIS} $ relates these incoming and outgoing states in a NIS junction, given as,}
{\begin{equation}
    S_{NIN} = \begin{pmatrix}
s_{11}^{ee}  & s_{12}^{ee} \\
s_{21}^{ee} & s_{22}^{ee}
\end{pmatrix} = \begin{pmatrix}
-i r  & t \\
t & -i r
\end{pmatrix}, 
\end{equation}}

{where, we consider $r$ to be the reflection amplitude of the electron to reflect as electron due to the insulating barrier, whereas $t$ is the transmission amplitude for the electron (hole) to transmit via the insulating barrier in NIN junction.}
{The reflection probability is $\mathcal{R}_{NIN} = |s_{11}^{ee}|^2$ and transmission probability is $\mathcal{T}_{NIN} = |s_{11}^{he}|^2$ in the NIN junction.}

{The general expression for quantum noise correlation in NIN junction is given as ~\cite{noise, thermalnoise}
\begin{eqnarray}
    Q^{NIN}_{pq} &=& \dfrac{2e^2}{h} \sum_{ \{k, l \} \in \{1,2\}} \int^{\infty}_{-\infty} A_{k l}(p,E) A_{l k }(q,E) \left( f_{\gamma e} (1- f_{\delta e}) + f_{\delta e} (1- f_{\gamma e})\right) dE
\end{eqnarray}}

{where $A_{k l }(p x,E) = \delta_{p k} \delta_{p l} - s_{p k}(E) s_{p l}(E)$.}

{Quantum noise auto-correlation ($\alpha = \beta = 1 $) in a NIN junction, can then be written as,
\begin{eqnarray}
Q^{NIN}_{11} &=& \dfrac{2e^2}{h} \sum_{ \{\gamma, \delta \} \in \{1,2\}} \int^{\infty}_{-\infty} A_{\delta \gamma}(1,E) A_{\gamma \delta }(1,E) \left( f_{\gamma e} (1- f_{\delta e}) + f_{\delta e} (1- f_{\gamma e}) \right) dE
\end{eqnarray}}

{Now, using the $s$-matrix $S_{NIN}$ as in Eq. (B18), we get}

{\begin{eqnarray}
Q_{11}^{NIN} &=& \dfrac{4e^2}{h} \int^{\infty}_{-\infty} \bigl\{  A_{11}(1,E) A_{11}(1,E) f_{1e} (1 - f_{1e}) 
+ A_{12} (1,E) A_{21}(1,E) f_{1e} (1 - f_{2e})  \nonumber \\
&& ~ + A_{21}(1,E) A_{12}(1,E) f_{2e} (1 - f_{1e}) + A_{22}(1,E) A_{22}(1,E) f_{2e} (1 - f_{2e}) \bigr\} dE \nonumber \\
&=& \dfrac{4e^2}{h} \int^{\infty}_{-\infty} \mathcal{T}_{NIN}^2 \{ f_{1e}(1-f_{1e})+f_{2e}(1-f_{2e}) \} + \mathcal{T}_{NIN}( 1 - \mathcal{T}_{NIN}) \{ f_{1e}(1-f_{2e})  + f_{2e} ( 1 -f_{1e}) \} dE \nonumber\\
&=& \dfrac{4e^2}{h} \int^{\infty}_{-\infty} \mathcal{T}_{NIN}^2 \{ f_{1e}(1-f_{1e})+f_{2e}(1-f_{2e}) \} + \mathcal{T}_{NIN}( 1 - \mathcal{T}_{NIN}) \{ f_{1e} - f_{1e}^2 + f_{1e}^2 - f_{1e} f_{2e}  + f_{2e} - f_{2e}^2 + f_{2e}^2 - f_{1e} f_{2e} \} dE \nonumber \\
&=& \dfrac{4e^2}{h} \int^{\infty}_{-\infty} \mathcal{T}_{NIN} (f_{1e}(1-f_{1e})+f_{2e}(1-f_{2e})) + \mathcal{T}_{NIN}( 1 - \mathcal{T}_{NIN}) (f_{1e}-f_{2e})^2 dE,  \nonumber \\ 
&=& \dfrac{4e^2}{h} \int^{\infty}_{-\infty} \left[ F^{NIN}_{11th} \{ f_{1e}(1-f_{1e})+f_{2e}(1-f_{2e}) \} + F^{NIN}_{11sh} (f_{1e}-f_{2e})^2 \right] dE,
\label{c1:eqn_s11_NIN}
\end{eqnarray}
where $F^{NIN}_{11sh}$ and $F^{NIN}_{11th}$ denote scattering terms for shot noise ($Q^{NIN}_{11sh}$) and thermal noise ($S^{NIN}_{11th}$) contributions which depend on reflection and transmission probabilities, i.e., $F^{NIN}_{11th} = \mathcal{T}_{NIN}$, and $F^{NIN}_{11sh} = \mathcal{T}_{NIN}( 1 - \mathcal{T}_{NIN}) $.  
The first term in Eq. (\ref{c1:eqn_s11_NIN}) represents the thermal noise contribution to total quantum noise. In contrast, the term with $(f_{1e}-f_{2e})^2$ in Eq. (\ref{c1:eqn_s11_NIN}) represents shot noise contribution to total quantum noise. The total quantum noise in a NIN junction is $Q^{NIN}_{11}=Q^{NIN}_{11th}+Q^{NIN}_{11sh}$, where the thermal noise-like contribution ($Q^{NIN}_{11th}$) and shot noise-like contribution ($Q^{NIN}_{11sh}$) are then given as,
\begin{eqnarray}
Q^{NIN}_{11th} &=& \dfrac{4e^2}{h} \int^{\infty}_{-\infty} F^{NIN}_{11th} (f_{1e}(1-f_{1e})+f_{2e}(1-f_{2e}))dE, \,\,\, \text{and}\,\,\, 
Q^{NIN}_{11sh} = \dfrac{4e^2}{h} \int^{\infty}_{-\infty} F^{NIN}_{11sh} (f_{1e}-f_{2e})^2~dE. 
\label{c1:eqn_Sshth_NIN}
\end{eqnarray}
}

\section{$\Delta_T$ noise in NIN junction}
\label{App_ch_Dn}

{To calculate the $\Delta_T$ noise in the NIN junction, i.e., $\Delta_T^{NIN}$, we use $Q_{11sh}^{NIN}$ from Eq. (\ref{c1:eqn_Sshth_NIN}).} {We define the temperature of the normal metal
as $T_1 = \bar{T} + \frac{\Delta T}{2}$ and the temperature of the superconductor as $T_2 = \bar{T} - \frac{\Delta T}{2}$ at zero voltage bias ($V = 0$), where $\bar{T}$ represents the average temperature of
the system and $\Delta T$ denotes the temperature bias. In the linear response regime, $\Delta T \ll \bar{T}$. From Eq. (\ref{c1:eqn_Sshth_NIN}), the $\Delta_T^{NIN}$ noise is given as,}
{
\begin{equation}
    \Delta_T^{NIN} = \frac{4e^2}{h} F_{11sh}^{NIN}  \int_{-\infty}^{\infty} dE  (f_{1e} - f_{2e})^2.
\label{eq15}    
\end{equation}}

 {where, $F_{11sh}^{NIN} = \mathcal{T}_{NIN} (1 - \mathcal{T}_{NIN})$. Now, the integration in Eq. (\ref{eq15}) can be done using the Taylor series expansion of $f_{1e}$ and $f_{2e}$ following Refs. \cite{atomicscaleexpt, dTtheory}. The Taylor series expansion of $f_{1e}$ and $f_{2e}$ {in the limit $\Delta T \ll \bar{T}$} around $T_1 = T_2 = \bar{T}$ up to first order in $\Delta T$ is given as,}
{
\begin{equation}
    f_{1e} = f - \frac{\Delta T}{2} \frac{\partial f}{\partial \bar{T}} \quad \text{and} \quad f_{2e} = f + \frac{\Delta T}{2} \frac{\partial f}{\partial \bar{T}},
\label{eq17}
\end{equation}}

{where, $f = \frac{1}{1+e^{\frac{E}{k_B \bar{T}}}}$ is the Fermi-Dirac distribution at temperature $\bar{T}$. Substituting Eq. (\ref{eq17}) to Eq. (\ref{eq15}), we get}
{
\begin{align}
\begin{split}
    \Delta_T^{NIN} &= \frac{4e^2}{h} F_{11sh}^{NIN} \int_{-\infty}^{\infty} dE \left(\Delta T \frac{\partial f}{\partial \bar{T}}\right)^2 = \frac{4e^2}{h}F_{11sh}^{NIN} \frac{(\Delta T)^2}{k_B^2 \bar{T}^4}  \int_{-\infty}^{\infty} dE  \frac{E^2 e^{\frac{2E}{k_B \bar{T}}}}{\left(1 + e^{\frac{E}{k_B \bar{T}}}\right)^4}.
    \label{eq201}
    \end{split}
\end{align}}

{
Using the relation: $
\int_{-\infty}^{\infty} dE \, \frac{E^2 \, e^{\frac{2E}{k_B \bar{T}}}}{\left(1 + e^{\frac{E}{k_B \bar{T}}}\right)^4} = \left(k_B \bar{T}\right)^3 \left(\frac{\pi^2}{18} - \frac{1}{3}\right),
$ we get $\Delta_T^{NIN}$ to be,}
{
\begin{equation}
    \Delta_T^{NIN} = \frac{4e^2}{h} F_{11sh}^{NIN}k_B \bar{T} \left(\frac{\pi^2}{18} - \frac{1}{3}\right) \frac{\Delta T^2}{\bar{T}^2}.
\label{eq16}
\end{equation}}

{Eq. (\ref{eq16}) is the exact expression for $\Delta_T$ noise in NIN junction at zero voltage bias and finite temperature bias.}

{
\section{Analytical derivation of ratio of $\Delta_T$ noise in transparent limit}
\label{ratio}
we consider the linear response regime where $\Delta T \ll \bar{T}$, where the temperature of the normal metal
is $T_1 = \bar{T} + \frac{\Delta T}{2}$ and the temperature of the superconductor is $T_2 = \bar{T} - \frac{\Delta T}{2}$. The expression of $\Delta_T^{NIN}$ and $\Delta_T^{NIS}$ at zero bias voltage (see, Eqs. (\ref{eq300}) and (\ref{eq101})) are given as
\begin{equation}
\Delta_T^{NIN} = \frac{4e^2}{h} F_{11sh}^{NIN}k_B \bar{T} \left(\frac{\pi^2}{18} - \frac{1}{3}\right) \frac{\Delta T^2}{\bar{T}^2}, \quad
\Delta_T^{NIS} = \frac{4e^2}{h} \frac{(\Delta T)^2}{k_B^2 \bar{T}^4}  \int_{-\infty}^{\infty} dE F_{11sh}^{NIS} \frac{E^2 e^{\frac{2E}{k_B \bar{T}}}}{\left(1 + e^{\frac{E}{k_B \bar{T}}}\right)^4},
\end{equation}
where, $F_{11sh}^{NIN} = \mathcal{T}_{NIN} (1 - \mathcal{T}_{NIN})$ and $F_{11sh}^{NIS} = \mathcal{A} (1 - \mathcal{A}) + \mathcal{B} (1 - \mathcal{B}) + 2 \mathcal{A} \mathcal{B}$. Here, $\mathcal{T}_{NIN}$ is the transmission probability for the NIN junction, whereas $\mathcal{A}$ and $\mathcal{B}$ are the Andreev reflection and normal reflection probabilities in NIS junction.
Calculation of the ratio $\frac{\Delta_T^{NIS}}{\Delta_T^{NIN}}$ for the transparent limit ($Z \to 0$) gives $(\frac{\Delta_T^{NIS}}{\Delta_T^{NIN}})_{Z \to 0} = 16$, whereas with increase in barrier strength ($Z$), the ratio decreases, such as $(\frac{\Delta_T^{NIS}}{\Delta_T^{NIN}})_{Z \to 1} \simeq 2$ and $(\frac{\Delta_T^{NIS}}{\Delta_T^{NIN}})_{Z \to 5} \simeq 0$. \\ 

Here in below, an analytical calculation of the ratio $\frac{\Delta_T^{NIS}}{\Delta_T^{NIN}}$ in the transparent limit is shown. First of all, the expression for $F_{11sh}^{NIN}$ is $
F_{11sh}^{\text{NIN}}=\mathcal T_{NIN}(1-\mathcal T_{NIN})=\frac{Z^2}{(1+Z^2)^2},$ where $\mathcal{T}_{NIN} = \frac{1}{ 1 + Z^2}$. In a NIS junction, the scattering probabilities for $ E < \Delta$, are 
$\mathcal{A} =\frac{\Delta^2}{E^2 + (\Delta^2 - E^2)(1 + 2Z^2)^2}$, \text{and} $\mathcal{B}=1-\mathcal{A}$ \cite{BTK}. This leads to,
\begin{align}
    F_{11sh}^{NIS}&=\mathcal{A}(1-\mathcal{A})+\mathcal{B}(1-\mathcal{B})+2\mathcal{A}\mathcal{B} =4\mathcal{A}\mathcal{B} = \frac{16(1-x^2)Z^2(1 + Z^2)}{(-1 + 4(-1+x^2)Z^2 + 4(-1 + x^2) Z^4)^2},
\end{align} where $x = \frac{E}{\Delta}$. In this manuscript, we have explicitly considered $k_B T \ll \Delta$, implying energies $E \ll \Delta$, i.e., $x \ll 1$. In this regime, $F_{11sh}^{NIS} = \frac{16 Z^2 (1 + Z^2)}{(1 + 2Z^2)^2}$. $F_{11sh}^{NIS}$ is thus energy independent.
In the expression for $\Delta_T^{NIS}$ as in Eq. (1), $F_{11sh}^{NIS}$ comes out of the energy integral and $\Delta_T^{NIS}$ in this limit is, 
\begin{equation}
    \Delta_T^{NIS} = \frac{4e^2}{h} \frac{(\Delta T)^2}{k_B^2 \bar{T}^4} F_{11sh}^{NIS} \int_{-\infty}^{\infty} dE \frac{E^2 e^{\frac{2E}{k_B \bar{T}}}}{\left(1 + e^{\frac{E}{k_B \bar{T}}}\right)^4},
\end{equation}
where $F_{11sh}^{NIS} = \frac{16 Z^2 (1 + Z^2)}{(1 + 2Z^2)^2}$.  As shown in Eq. (\ref{eq16}) of our paper, the integral $\int_{-\infty}^{\infty} dE \frac{E^2 e^{\frac{2E}{k_B \bar{T}}}}{\left(1 + e^{\frac{E}{k_B \bar{T}}}\right)^4}$ is $(k_B \bar{T})^3 \left(\frac{\pi^2}{18} - \frac{1}{3}\right) \frac{\Delta T^2}{\bar{T}^2}$. Thus, 
\begin{equation}
    \Delta_T^{NIS}=\frac{4e^2}{h} F_{11sh}^{NIS} k_B \bar{T} \left(\frac{\pi^2}{18} - \frac{1}{3}\right) \frac{\Delta T^2}{\bar{T}^2},\,\,\text{and ratio:}\,\, \frac{\Delta_T^{NIS}}{\Delta_T^{NIN}}=\frac{F_{11sh}^{NIS}}{F_{11sh}^{NIN}}=\frac{16(1+Z^2)^3}{(1+2Z^2)^2},
\end{equation}
which in limit $Z \to 0$ reduces to $\frac{\Delta_T^{NIS}}{\Delta_T^{NIN}} = 16.$

% \begin{equation}
%     \frac{\Delta_T^{NIS}}{\Delta_T^{NIN}} = 16
% \end{equation}

 Similarly the ratio for normalised $\Delta_T$ noise $\frac{\bar{\Delta}_T^{NIS}}{\bar{\Delta}_T^{NIN}}$ can be found, as $\bar{\Delta}_T^{NIS}=\frac{\Delta_T^{NIS}}{G_{NIS}}$, and $\bar{\Delta}_T^{NIN}=\frac{\Delta_T^{NIN}}{G_{NIN}}$. In the transparent limit $\frac{G_{NIN}}{G_{NIS}}=\frac{1}{2}$, hence $\frac{\bar{\Delta}_T^{NIS}}{\bar{\Delta}_T^{NIN}}=\frac{1}{2}\frac{\Delta_T^{NIS}}{\Delta_T^{NIN}}=8.$
}

\end{widetext}


\begin{thebibliography}{}

%\bibitem{qnoise}
%C. Beenakker and C. Sch\"{o}nenberger, Quantum Shot Noise, Phys. Today 56, 5, 37 (2003).


%\bibitem{shotnoise}
%M. Henny, et. al., The Fermionic Hanbury Brown and Twiss Experiment, Science 284, 296 (1999); T. Mohapatra, S. Pal, and C. Benjamin, Probing the topological character of superconductors via nonlocal Hanbury Brown and Twiss correlations, Phys. Rev. B 106, 125402 (2022).

\bibitem{atomicscaleexpt}
O. S. Lumbroso, et. al., Electronic noise due to temperature differences in atomic-scale junctions, Nature 562, 240 (2018).

\bibitem{qcircuitexpt}
E. Sivre, et. al., Electronic heat flow and thermal shot noise in quantum circuits, Nat. Comm. 10, 5638 (2019).

\bibitem{tjunctionexpt}
S. Larocque, et. al., Shot Noise of a Temperature-Biased Tunnel Junction, Phys. Rev. Lett. 125, 106801 (2020).

% \bibitem{suppl}
% See Supplemental Material at " " for Mathematica code to plot the thermovoltage and $\Delta_T$ noise.

\bibitem{DeltaNS}
E. Zhitlukhina, M. Belogolovskii, P, Seidel, Electronic noise generated by a temperature gradient across a hybrid normal metal–superconductor nanojunction, Applied Nanosci. 10, 5121 (2020).

\bibitem{dTtheory}
J. Rech, et. al., Negative Delta-T Noise in the Fractional Quantum Hall Effect, Phys. Rev. Lett. 125, 086801 (2020).

\bibitem{popoff}
A. Popoff, et. al., Scattering theory of non-equilibrium noise and delta T current fluctuations through a quantum dot, J. Phys.: Condens. Matter 34, 185301 (2022).

\bibitem{generalbound}
J. Eriksson, et. al., General Bounds on Electronic Shot Noise in the Absence of Currents, Phys. Rev. Lett. 127, 136801 (2021);
L. Tesser, et. al., Charge, spin, and heat shot noises in the absence of average currents: Conditions on bounds at zero and finite frequencies, Phys. Rev. B 107, 075409 (2023).

\bibitem{machinelearn}
Gerry, et. al., Machine learning delta-T noise for temperature bias estimation,  J. Chem. Phys.
162, 084108 (2025).

\bibitem{Prokudina2024}
M. G. Prokudina, A. F. Shevchun, and E. S. Tikhonov, 
``Local thermometry of NbSe$_2$ flake with delta-T noise measurements,'' 
arXiv:2407.09075 (2024).

% \bibitem{deltaNsfN}
 % T. Mohapatra and C. Benjamin, Spin-flip scattering engendered negative $\Delta_T$ noise, arXiv:2307.14072 (2023).

% \bibitem{qstat}
% G. Zhang, I. V. Gornyi and C. Spaanslatt, Delta-T noise for weak tunneling in one-dimensional systems: Interactions versus quantum statistics, Phys. Rev. B 105, 195423 (2022).


%\bibitem{QPCbuttiker}
%M. Buttiker, Quantized transmission of a saddle-point constriction, Phys. Rev. B 41, 7906 (1990).

%\bibitem{beneti}
%G. Beneti, et. al., Fundamental aspects of steady-state conversion of heat to work at the nanoscale, Phys. Rep. 694, 1 (2017).

%\bibitem{Houten}
%H. van Houten, Thermo-electric properties of quantum point contacts, Semicond. Sci. Technol. 7, B215 (1992).

%\bibitem{Sdatta} 
%S. Datta, \text{Electronic transport in mesoscopic systems}, 
%Cambridge University Press, (1997).

%\bibitem{QPCwees}
%B. J. van Wees, et. al., Quantized conductance of point contacts in a two-dimensional electron gas, Phys. Rev. Lett 60, 848 (1988).

%\bibitem{QPCkamp}
%L. W. Molenkamp, et. al., Peltier coefficient and thermal conductance of a quantum point contact, Phys. Rev. Lett. 68, 3765 (1992).

%\bibitem{QPCwharam}
%D. A. Wharam, et. al., One-dimensional transport and the quantisation of the ballistic resistance, J. Phys. C: Solid State Phys. 21, L209 (1988).

%\bibitem{heatcurrent}
%Y. Dubi and M. Di Ventra, Heat flow and thermoelectricity in atomic and molecular junctions, Rev. of Mod. Phy. 83, 131 (2011).

%\bibitem{shotNSN}
%R. Mélin, C. Benjamin, and T. Martin, 
%Positive cross-correlations of noise in superconducting hybrid structures: Roles of interfaces and interactions, Phys. Rev. B 77, 094512 (2008).

%\bibitem{NQS}
%H. T. Man, T. M. Klapwijk, and A.F. Morpurgo, Transport through a superconductor-interacting normal metal junction: a phenomenological description, arXiv:050456v16 (2005). 

\bibitem{BTK}
G. E. Blonder, M. Tinkham, and T. M. Klapwijk, Transition from metallic to tunneling regimes in superconducting microconstrictions: Excess current, charge imbalance, and supercurrent conversion, Phys. Rev. B 25, 4515 (1982).

\bibitem{Benenti}
G. Benenti, G. Casati, K. Saito, and R. Whitney, Fundamental aspects of steady-state conversion of
heat to work at the nanoscale, Phys. Rep. 694, 1 (2017).

\bibitem{Tc}
Piotr Magierski, Gabriel Wlazlowski, and Aurel Bulgac, Onset of a Pseudogap Regime in Ultracold Fermi Gases,  Phys. Rev. Lett. 107, 145304 (2011). 

\bibitem{datta}
M. P. Anantram and S. Datta, Current fluctuations in mesoscopic systems with Andreev scattering, Phys. Rev. B 53, 16390 (1996).



\bibitem{noise}
Ya M. Blanter, and M. B\"{u}ttiker, Shot Noise in Mesoscopic Conductors, Phys. Reports 336, 1 (2000).

\bibitem{thermalnoise}
L. P. Kouwenhoven, G. Sch\"{o}n, and L.L. Sohn, Introduction to mesoscopic electron transport, NATO ASI Series E, Kluwer Academic Publishing, Dordrecht, 225, 345 (1997).


\bibitem{martin}
T. Martin, Noise in mesoscopic physics, Les Houches Session LXXXI, H. Bouchiat et. al. eds. (Elsevier 2005).




\bibitem{github} 
Mathematica notebook for the thermovoltage and $\Delta_{T}$ noise plots can be found in \href{https://github.com/Sachiraj/Delta-T-NIN-and-NIS.git}{$\Delta^{NIS}_{T}$ noise code}.

\bibitem{Buttiker1993}
M. Büttiker, Capacitance, admittance, and rectification properties of small conductors, 
{J. Phys.: Condens. Matter} \textbf{5}, 9361 (1993); 
T. Christen and M. Büttiker, Gauge-invariant nonlinear electric transport in mesoscopic conductors, 
{Europhys. Lett.} \textbf{35}, 523 (1996); 
D. Sánchez and R. López, Scattering theory of nonlinear thermoelectric transport, 
{Phys. Rev. Lett.} \textbf{110}, 026804 (2013).

\bibitem{Lopez2013}
R. López and D. Sánchez, Nonlinear heat transport in mesoscopic conductors: Rectification, Peltier effect, and Wiedemann-Franz law, 
{Phys. Rev. B} \textbf{88}, 045129 (2013); 
J. Meair and P. Jacquod, Scattering theory of nonlinear thermoelectricity in quantum coherent conductors, 
{J. Phys.: Condens. Matter} \textbf{25}, 082201 (2013).

\bibitem{Tc1}
Ø. Fischer et al., Scanning tunneling spectroscopy of high-temperature superconductors, Rev. Mod. Phys. 79, 353 (2007)


\bibitem{heatcurrentexpt}
S. Jezouin, et. al., Quantum Limit of Heat Flow Across a Single Electronic Channel, Science 342, 601 (2013); D.E. Angelescu, M.C. Cross and M.L. Roukes, Heat Transport in Mesoscopic Systems, Superlattices and Microstructures 23, 673 (1998).

\bibitem{NIS}
Jiasen Niu, et. al., Why Shot Noise Does Not Generally Detect Pairing in Mesoscopic Superconducting Tunnel Junctions, Phys. Rev. Lett. 132, 076001 (2024).

\bibitem{Fano1}
B. T. Matthias, T. H. Geballe, S. Geller, and E. Corenzwit, Superconductivity of Nb$_3$Sn, Phys. Rev. 95, 1435 (1954).

\bibitem{Fano}
Z. Sun, Z. Baraissov, R. D. Porter, L. Shpani, Y.-T. Shao, T.
Oseroff, M. O. Thompson, D. A. Muller, and M. U. Liepe, high-purity Nb$_3$Sn
superconducting RF resonant cavity by seed-free
electrochemical synthesis, Supercond. Sci. Technol. 36, 115003 (2023).

\bibitem{blonder}G. E. Blonder, M. Tinkham, Metallic to tunneling transition in Cu-Nb point contacts, {Phys. Rev. B} \textbf{27}, 112 (1983).

\bibitem{lambert}
C. J. Lambert, Generalized Landauer formulae for quasiparticle transport in disordered superconductors, J. Phys.: Condens. Matter 3, 6579 (1991).







%\bibitem{heatnoise}
%P. Eymeoud and A. Crepieux, Mixed electrical-heat noise spectrum in a quantum dot, Phys. Rev. B 94, 205416 (2016); A. Crepieux and F. Michelini, Mixed, charge and heat noises in thermoelectric nanosystems, J. Phys.: Condens. Matter 27, 015302 (2015).



% \bibitem{kondo}
% Delta-T noise in the Kondo regime, M. Hasegawa and K. Saito, Phys. Rev. B 103, 045409 (2021).

% \bibitem{zerocurrent} 
% G. Benenti, G. Casati, K. Saito, and R. S. Whitney,
% Fundamental aspects of steady-state conversion of heat to
% work at the nanoscale, Phys. Rep. 694, 1 (2017); F. Binder, L. A. Correa, C. Gogolin, J. Anders, and G. Adesso, Thermodynamics in the Quantum Regime (Springer International Publishing, Cham, Switzerland, 2018).

% \bibitem{zerocurrentnoise}
% R. Sánchez, J. Splettstoesser, and R. S. Whitney, Non-equilibrium System as a Demon, Phys. Rev. Lett. 123, 216801 (2019); S. E. Deghi and R. A. Bustos-Marún, Entropy current and efficiency of quantum machines driven by non-equilibrium incoherent reservoirs, Phys. Rev. B 102, 045415 (2020).

% \bibitem{zerocurrentnoise1}
% R. Sánchez, B. Sothmann, A. N. Jordan, and M. Büttiker, Correlations of heat and charge currents in quantum-dot thermoelectric engines, New J. Phys. 15, 125001 (2013); N. Freitas and M. Esposito, Characterizing autonomous Maxwell demons, Phys. Rev. E 103, 032118 (2021).


% \bibitem{nonlinear}
% G. Marchegini, et. al., Nonlinear Thermoelectricity with Electron-Hole Symmetric systems, Phys. Rev. Lett. 124, 106801 (2020)

% \bibitem{Hirsch}
% F. Marsiglio and J. E. Hirsch, Hole superconductivity and the high-T$_c$, oxides,  Phys. Rev. B 41, 6435 (1990).



\end{thebibliography}
\end{document}